\documentclass[10pt]{article}
\usepackage{latexsym,amsmath,amssymb,graphics,amscd}
\usepackage{multirow}
\usepackage{booktabs}
\usepackage{tabularx}
\usepackage[hang,footnotesize]{caption}
\usepackage[pdftex]{graphicx}

\addtocounter{footnote}{1}
\makeatletter
\@addtoreset{figure}{section}
\def\thefigure{\thesection.\@arabic\c@figure}
\def\fps@figure{h,t}
\@addtoreset{table}{bsection}
\def\thetable{\thesection.\@arabic\c@table}
\def\fps@table{h, t}
\@addtoreset{equation}{section}

\makeatother

\newtheorem{theorem}{Theorem}[section]
\newtheorem{definition}[theorem]{Definition}
\newtheorem{lemma}[theorem]{Lemma}
\newtheorem{remark}[theorem]{Remark}
\newtheorem{proposition}[theorem]{Proposition}

\newsavebox{\savepar}

\makeatletter

\begin{document}

\title{\textbf{Multivariate GARCH estimation via a Bregman-proximal trust-region method}} 
\author{St\'ephane Chr\'etien$^{1}$ and Juan-Pablo Ortega$^{2}$}
\date{}
\maketitle

\begin{abstract}
The estimation of multivariate GARCH time series models is a difficult task mainly due to the significant overparameterization exhibited  by the problem and usually referred to as the ``curse of dimensionality". For example, in the case of the VEC family, the number of 
parameters involved in the model grows as a polynomial of order four on the dimensionality of the problem. Moreover, these parameters are subjected to convoluted nonlinear constraints necessary to ensure, for instance, the existence of stationary solutions and the positive semidefinite character of the conditional covariance matrices used in the model design. So far, this problem has been addressed in the literature only in low dimensional cases  with strong parsimony constraints (see for instance~\cite{altay:pinar:leyffer} for the diagonal three-dimensional {\rm VEC} handled with ad-hoc techniques). In this paper we propose a general formulation of the 
estimation problem in any dimension and develop a Bregman-proximal trust-region method for its solution. The Bregman-proximal 
approach allows us to handle the constraints in a very efficient and natural way by staying in the primal space and the Trust-Region mechanism 
stabilizes and speeds up the scheme. Preliminary computational experiments are presented and confirm the very good performances of the proposed approach.
\end{abstract}

\bigskip


\makeatletter
\addtocounter{footnote}{1} \footnotetext{%
D\'{e}partement de Math\'{e}matiques de Besan\c{c}on, , Probability and Statistics Group, Universit\'{e} de Franche-Comt\'{e}, UFR des
Sciences et Techniques. 16, route de Gray. F-25030 Besan\c{c}on cedex.
France. {\texttt{Stephane.Chretien@univ-fcomte.fr} }}
\addtocounter{footnote}{1} \footnotetext{%
Centre National de la Recherche Scientifique, D\'{e}partement de Math\'{e}matiques de Besan\c{c}on, , Probability and Statistics Group, Universit\'{e} de Franche-Comt\'{e}, UFR des
Sciences et Techniques. 16, route de Gray. F-25030 Besan\c{c}on cedex.
France. {\texttt{Juan-Pablo.Ortega@univ-fcomte.fr} }}
\makeatother

\section{Introduction}

Autoregressive conditionally heteroscedastic (ARCH) models~\cite{engle:arch} and their generalized counterparts (GARCH)~\cite{bollerslev:garch} are standard econometric tools to capture the leptokurticity and the volatility clustering exhibited by financial time series. In the one dimensional situation, a large collection of parametric models that account for various stylized features of financial returns is available. Additionally,  adequate model selection and estimation tools have been developed, as well as explicit characterizations of the conditions that ensure stationarity or the existence of higher moments. 

One of the advantages of GARCH models that makes them particularly useful is that once they have been calibrated they provide an estimate of the dynamical behavior of volatility which, in principle, is not directly observable. This feature makes desirable the extension of the GARCH prescription to the multivariate case since such a generalization provides a dynamical picture of the correlations between different assets which are of major importance, in the context of financial econometrics, for pricing and hedging purposes, asset allocation, and risk management in general. 

This generalization is nevertheless not free from difficulties. The most general multivariate GARCH models are the VEC prescription proposed by Bollerslev {\it et al}~\cite{bollerslev:vec} and the BEKK model by Engle {\it  et al}~\cite{engle:bekk}; both families of models present satisfactory properties that match those found in univariate GARCH models, nevertheless their lack of parsimony, even in low dimensions makes them extremely difficult to calibrate; for example, VEC(1,1) models require $n(n+1)(n(n+1)+1)/2  $ parameters, where $n$ is the dimensionality of the modeling problem; BEKK(1,1,1) requires $n(5n+1)/2 $. Indeed, due to the high number of parameters needed, it is rare to find these models at work beyond two or three dimensions and even then, ad hoc estimation techniques are used and additional limitations are imposed on the model to make it artificially parsimonious; see for example~\cite{altay:pinar:leyffer} for an illustration of the estimation of a three dimensional DVEC model (VEC  model with diagonal parameter matrices~\cite{bollerslev:vec}) using constrained non-linear programming. These difficulties have lead to the search for more parsimonious but still functioning models like for example CCC~\cite{bollerslev:ccc}, DCC~\cite{tse:dcc, engle:dcc} or GDC~\cite{kroner:ng}. On a different vein, a number of  different signal separation techniques have been tried out in the financial time series context with the aim of reducing this intrinsically multivariate problem to a collection of univariate ones. For example, principal component analysis is used in the O-GARCH model~\cite{ding:thesis, alexander:chibumba, alexander:ogarch, alexander:covariance_matrices} and independent component analysis in the  ICA-GARCH model~\cite{wu:PCA, pena:ica}. We advice the reader to check with the excellent reviews~\cite{bauwens:garch_review, review:multivariate:GARCH:handbook} for a comprehensive description of these and other models.

Despite the overparameterization problem we will concentrate in this work on full fledge VEC models. This decision is taken not for the pure sake of generality but because the intrinsic difficulties of this parametric family of models make them an ideal benchmark for testing optimization techniques subjected to potentially complex matrix constraints. Stated differently, it is our belief that, independently from the pertinence of the VEC family in certain modeling situations, any optimization algorithm developed to estimate them will work smoothly when applied to more elementary situations. Hence, the work that we present in this paper is capable of increasing the range of dimensions in which VEC models can be estimated in practice by improving the existing technology in two directions:
\begin{itemize}
\item Explicit matrix formulation of the model and of the associated stationarity and positivity constraints: the works in the literature usually proceed by expressing  the constraints in terms of the entries of the parameter matrices (see for example~\cite{altay:pinar:leyffer} in the DVEC case). A global matrix formulation is necessary in order to obtain a dimension independent encoding of the problem. This task is carried out in Sections~\ref{Preliminaries on matrices and matrix operators} and~\ref{The VEC-GARCH model}
\item Use of a Bregman-type proximal optimization algorithm that efficiently handles the constraints in the primal space. More specifically, we will be using {\bf Burg's matrix divergence}; this divergence is  presented, for example, in~\cite{kulis:sustik:dhillon} and it is a particular instance of a {\bf Bregman divergence}. Bregman divergences are of much use in the context of machine learning (see for instance~\cite{dhillon:tropp, kulis:sra:dhillon} and references therein). In our situation we have opted for this technique as it allows for a particularly efficient treatment of positive definiteness constraints, as those in our problem, avoiding the need to solve additional secondary optimization problems that appear, for example, had we used Lagrange duality. It is worth emphasizing that even though the constraints that we handle in the estimation problem admit a simple and explicit conic formulation well adapted to the use of Lagrange multipliers, the associated dual optimization problem is in this case of difficulty comparable to that of the primal so avoiding this extra step is a major advantage. This approach is presented in Sections~\ref{Constrained optimization via Bregman divergences} and~\ref{Bregman divergences for VEC models}. In Section~\ref{Performance improvement: BFGS and trust-region corrections} we couple the use of Bregman divergences with a refinement of the local penalized  model using  quadratic BFGS type terms and with a trust-region iteration acceptance rule that greatly stabilizes the primal trajectory and improves the convergence speed of the algorithm. Finally, given the non-linear non-convex nature of estimation via quasi-loglikelihood optimization, the availability of good preliminary estimation techniques is of paramount importance in order to avoid local minima; this point is treated in Section~\ref{Preliminary estimation} where some of the simpler modeling solutions listed above are used to come up with a starting point to properly initialize the optimization algorithm.  
\end{itemize}

In Section~\ref{numerical experiments} we illustrate the estimation method proposed in Section~\ref{Calibration via Bregman matrix divergences}  with various numerical experiments that prove its applicability and support the following statements:
\begin{itemize}
\item The trust-region correction speeds up the algorithm and the BFGS modification makes the convergence rate dimensionally independent.
\item More importantly, VEC seems to be a performing modeling tool for stock market log-returns when compared with other more parsimonious parametric families, {\it  even in dimensions where the high number of parameters in comparison with the sample size would make us expect a deficient modeling behavior}. Our conjecture is that this better than expected results have to do with the spectral sparsity (in the dimensions we work on we should rather say spectral concentration) of the correlation matrices exhibited by  stock market log-returns; this empirically observed feature imposes nonlinear constraints on the model parameters that invalidate the hypotheses necessary to formulate the standard results on the asymptotic normality of the quasi-loglikelihood parameter estimator (see later on expressions~(\ref{estimator properties 1}) and (\ref{estimator properties 2})) and make it more favorable with respect to its use with standard sample sizes. In a forthcoming publication we plan to provide a detailed study of the convergence and complexity properties of the proposed algorithm, together with dimension reduction techniques based on the use of the spectral sparsity that, as we said, is empirically observed in actual financial time series.
\end{itemize}

\medskip

\noindent {\bf Notation and conventions:} In order to make the reading of the paper easier, most   of the proofs of the stated results have been gathered at the end in the form of appendices. All along the paper, bold symbols like $\mathbf{r} $ denote column vectors, $ \mathbf{r} ^T  $ denotes the transposed vector.
Given a filtered probability space $(\Omega, \mathbb{P}, \mathcal{F}, \{ \mathcal{F} _t\}_{t \in \mathbb{N}})$ and $X,Y $ two random variables, we will denote by $E _t[X]:=E[X| \mathcal{F}  _t]$ the conditional expectation, ${\rm cov} _t(X,Y):= {\rm cov}(X,Y| \mathcal{F} _t):=E_t[XY]-E_t[X]E _t[Y]$ the conditional covariance, and by ${\rm var} _t(X):=E_t[X ^2]-E _t[X] ^2 $ the conditional variance. A discrete-time stochastic process $\{X _t\}_{t \in  \mathbb{N}} $ is predictable when $X _t $ is $\mathcal{F} _{t-1} $-measurable, for any $t \in \mathbb{N} $.

\medskip

\section{Preliminaries on matrices and matrix operators}
\label{Preliminaries on matrices and matrix operators}

\noindent {\bf Matrices:} Let $n,m \in \mathbb{N} $ and  denote by $\mathbb{M}_{n,m}$ the space of $n \times m $ matrices. When $n=m $ we will just write $\mathbb{M} _n$ to refer to the space of $n \times n $ square matrices. Unless specified otherwise, all the matrices in this paper will contain purely real entries. The equality $A =  \left(a_{ij} \right)$ denotes the matrix $A$ with components $a_{ij}\in \mathbb{R}$. The symbol $\Bbb S_n $ denotes the subspace of $\mathbb{M} _n $ that contains all symmetric matrices
\begin{equation*}
\Bbb S_n=\{ A \in \mathbb{M}_m\mid A ^T=A\}
\end{equation*}
and $\Bbb S_n^+ $ (respectively $\Bbb S_n^-$) is the cone in $\Bbb S _n $ containing the positive (respectively negative) semidefinite matrices. The symbol $A\succeq 0 $ (respectively $A\preceq 0 $) means that $A$ is positive (respectively negative) semidefinite. 

We will consider $\mathbb{M}_{n,m}$ as an inner product space with the pairing
\begin{equation}
\label{frobenius inner product}
\langle A,B\rangle= {\rm trace}(AB ^T)
\end{equation}
and denote by $\|A\|= \langle A,A\rangle^ \frac{1}{2}$ the associated Frobenius norm. Given a linear operator $\mathcal{A}:\mathbb{M}_{n,m} \rightarrow \mathbb{M}_{p,q} $ we will denote by $\mathcal{A} ^\ast : \mathbb{M}_{p,q} ^\ast  \rightarrow \mathbb{M}_{n,m}^\ast   $ its adjoint with respect to the inner product~(\ref{frobenius inner product}).

\medskip

\noindent {\bf The vec, vech, mat, and math operators and their adjoints:} The symbol ${\rm vec}:\mathbb{M}_n \rightarrow \mathbb{R}^{n ^2}$ denotes the operator that stacks all the columns of a matrix into a vector. Let $N= \frac{1}{2}n(n+1)$ and let ${\rm vech}: \mathbb{S}_n \rightarrow \mathbb{R}^{N}$ be the operator that stacks only the lower triangular part, including the diagonal, of a symmetric matrix into a vector. The inverse of the vech (respectively vec)operator will be denoted by ${\rm math}: \mathbb{R}^N \rightarrow \Bbb S_n $ (respectively ${\rm mat}: \mathbb{R}^{n^2} \rightarrow \Bbb M_n $).

Given $n \in \mathbb{N} $ and $N= \frac{1}{2}n(n+1)$, let $S=\{(i,j)\in \{1, \ldots, n\}\times \{1, \ldots, n\}\mid i\geq j\} $  we define $\sigma: S \rightarrow \{1, \ldots, N\}$ as the map that yields the position of component $(i,j), i\geq j,$ of any symmetric matrix in its equivalent vech representation. The symbol $\sigma^{-1}:\{1, \ldots, N\} \rightarrow S $ will denote its inverse and $\widetilde{\sigma}:\{1, \ldots, n\}\times \{1, \ldots, n\} \rightarrow  \{1, \ldots, N\}  $ the extension of $\sigma$ defined by:
\begin{equation}
\label{extension of sigma}
\widetilde{\sigma}(i,j)=\left\{
\begin{array}{ll}
\sigma(i,j) &i \geq j\\
\sigma(j,i) &i < j.
\end{array}
\right.
\end{equation}
The proof of the following result is provided in the Appendix.

\begin{proposition}
\label{vec identities}
Given $n \in \mathbb{N} $ and $N= \frac{1}{2}n(n+1)$, let $A \in \Bbb S _n $ and $m \in \mathbb{R}^N  $ arbitrary. The following identities hold true:
\begin{description}
\item [(i)]$\langle {\rm vech}(A), m\rangle= \frac{1}{2}\langle A+ {\rm diag}(A), {\rm math}(m) \rangle. $
\item [(ii)]$\langle A, {\rm math}(m)\rangle= 2\langle {\rm vech}(A- \frac{1}{2} {\rm diag}(A)), m \rangle,$
\end{description}
where ${\rm diag}(A) $ denotes the diagonal matrix obtained out of the diagonal entries of $A$.
Let ${\rm vech}^\ast :\mathbb{R}^N \rightarrow \Bbb S_n $ and $ {\rm math}^\ast :\Bbb S_n\rightarrow \mathbb{R}^N $ be the adjoint maps of ${\rm vech} $  and ${\rm math}$, respectively, then:
\begin{eqnarray}
{\rm math}^\ast(A)&=&2\, {\rm vech} \left( A- \frac{1}{2} {\rm diag}(A)\right), \label{expression maths}\\
{\rm vech}^\ast(m) &= &\frac{1}{2} \left( {\rm math}(m) +{\rm diag}({\rm math}(m))\right). \label{expression vechs}
\end{eqnarray}
The operator norms of the mappings that we just introduced are given by:
\begin{eqnarray}
\| {\rm vech}\|_{op} &=&1\label{operator norm vech}\\
\| {\rm math}\|_{op} &=&\sqrt{2}\label{operator norm math}\\
\| {\rm vech^\ast }\|_{op} &=&1\label{operator norm math2}\\
\| {\rm math}^\ast \|_{op} &=&\sqrt{2}\label{operator norm math3}\\
\| {\rm diag}\|_{op} &=&1\label{operator norm math4}
\end{eqnarray}
\end{proposition}

\medskip

\noindent {\bf Block matrices and the $\Sigma$ operator:} let $n \in \mathbb{N} $  and $B \in \mathbb{M}_{n ^2}$. The matrix $B$ can be divided into $n^2 $ blocks $B_{ij} \in \mathbb{M}_n  $ and hence its components can be labeled using a blockwise notation by referring to the $(k,l)$ element of the $(i,j)$ block as $(B_{ij})_{kl}$.  This notation makes particularly accessible the interpretation of $B$ as the coordinate expression of a linear endomorphism of the tensor product space $\mathbb{R}^n\otimes \mathbb{R}^n$. Indeed if $\{ \mathbf{e}_1, \ldots, \mathbf{e}_n \} $ is the canonical basis or $\mathbb{R} ^n $, we have
\begin{equation}
\label{components block matrices}
B (\mathbf{e}_i \otimes \mathbf{e}_k)=\sum_{j,l=1}^n (B_{ij})_{kl} (\mathbf{e}_j \otimes \mathbf{e}_l).
\end{equation} 

\begin{definition}
\label{sigma operator}
Let $A\in \Bbb M_N$ with $N= \frac{1}{2}n(n+1)$. We define $\Sigma (A) \in \Bbb S_{n ^2} $ blockwise using the expression
\begin{equation}
\label{sigma definition}
\left\{
\begin{array}{lcl}
\text{If $k\geq l $}&(\Sigma (A)_{kl})_{ij} &=\left\{
\begin{array}{cc}
\frac{1}{2}A_{\sigma(k,l), \sigma(i,j)},  &\text{if \quad}i>j\\
A_{\sigma(k,l), \sigma(i,j)},  &\text{if \quad}i=j\\
\frac{1}{2}A_{\sigma(k,l), \sigma(j,i)},  &\text{if \quad}i<j
\end{array}
\right.\\
 \ &\   & \\

\text{If $k\leq l $}&\Sigma (A)_{kl} &=\Sigma (A)_{lk}, 
\end{array}
\right.
\end{equation}
where $\sigma $ is the map defined above that yields the position of component $(i,j), i\geq j,$ of any symmetric matrix in its equivalent vech representation. By construction $(\Sigma (A)_{kl})_{ij}  $ is symmetric with respect to transpositions in the $(k,l)$ and $(i,j)$ indices; this implies that $\Sigma (A) $ is both symmetric and blockwise symmetric. We will refer to any matrix in $\Bbb S_{n ^2}$ with this property as {\bf $n$-symmetric} and will denote the corresponding space by  $\Bbb S_{n ^2}^n$.
\end{definition}

The proofs of the next two results are provided in the Appendix.

\begin{proposition}
\label{property of sigmaa statement}
Given $H \in \Bbb S _n $ and $A\in \Bbb M_N$, with $N= \frac{1}{2}n(n+1)$, the $n$-symmetric matrix $\Sigma(A)  \in \Bbb S_{n ^2}^n $ that we just defined satisfies:
\begin{equation}
\label{property of sigmaa}
A \, {\rm vech}(H)= {\rm vech}( \Sigma(A)\bullet H),
\end{equation}
where $\Sigma(A)\bullet H \in \Bbb S_n $ is the symmetric matrix given by
\begin{equation*}
(\Sigma(A)\bullet H)_{kl}= \langle\Sigma(A)_{kl},H\rangle= \operatorname{trace}(\Sigma (A)_{kl}H).
\end{equation*}
\end{proposition}

\begin{proposition}
\label{sigma dual}
Let $\Sigma: \mathbb{M}_N \rightarrow \Bbb M_{n^2}$ be the operator defined in the previous proposition, $N= \frac{1}{2}n(n+1)$. Then, for any $\mathcal{B} \in M_{n^2}$, the corresponding dual map $\Sigma^\ast : \Bbb M_{n^2} \rightarrow \mathbb{M}_N  $ is given by
\begin{equation}
\label{expression sigma dual}
\Sigma^\ast (\mathcal{B})=2B- \widetilde{B},
\end{equation}
where $B, \widetilde{B} \in \mathbb{M}_N  $ are the matrices defined by
\begin{equation*}
B_{pq}=((\mathbb{P}_{n ^2 }^n(\mathcal{B}))_{\sigma ^{-1}(p)})_{\sigma^{-1}(q)}, \quad\text{and} \quad \widetilde{B}_{pq}=B_{pq} \delta_{{\rm pr}_1(\sigma ^{-1}(p)),{\rm pr}_2(\sigma ^{-1}(p))}. 
\end{equation*}
The symbol $\mathbb{P}_{n ^2 }^n(\mathcal{B}) $ denotes the orthogonal projection of $\mathcal{B} \in \mathbb{M}_{n ^2} $ onto the space $\Bbb S_{n ^2 } ^n  $ of $n$-symmetric matrices that we spell out in Lemma~\ref{orthogonal projection n symmetric}. As we saw in Proposition~\ref{property of sigmaa statement},  $\Sigma $ maps into the space $\Bbb S_{n ^2 } ^n  $ of symmetric matrices; let $ \widetilde{\Sigma}:\mathbb{M}_N \rightarrow \Bbb S_{n^2}^n $ be the map obtained out of $\Sigma $ by restriction of its range. The map $\widetilde{\Sigma} $ is a bijection with inverse $\widetilde{\Sigma}^{-1}:\Bbb S_{n^2}^n \rightarrow \mathbb{M}_N $ given by
\begin{equation}
\label{sigma inverse expression}
\left(\widetilde{\Sigma}^{-1}(B)\right)_{p,q}= \left(B_{\sigma^{-1}(p)}\right)_{\sigma^{-1}(q)}\left(2-\delta_{{\rm pr}_1(\sigma ^{-1}(q)),{\rm pr}_2(\sigma ^{-1}(q))}\right).
\end{equation}
\end{proposition}

\section{The VEC-GARCH model}
\label{The VEC-GARCH model}

Consider the $n$-dimensional  conditionally heteroscedastic discrete-time process $\{  \mathbf{z}_t \} $ determined by the relation
\begin{equation*}
\mathbf{z}_t = H_t^{1/2} \boldsymbol{\epsilon _t} \quad\quad\text{with}\quad\quad
\{\boldsymbol{\epsilon _t}\}\sim {\rm IIDN}({\boldsymbol 0}, {\boldsymbol I}_n).
\end{equation*}
In this expression, $\{H_t\} $ denotes a predictable matrix process, that is for each $t \in \mathbb{N}  $, the matrix random variable $H_t $ is $\mathcal{F}_{t-1} $-measurable, and $H_t^{1/2} $ is a square root of  $H_t $, hence it satisfies $H_t^{1/2}(H_t^{1/2})^T=H_t $. In these conditions it is easy to show that the conditional mean $E_t[{\bf z} _t]=\boldsymbol{0}  $ and that the conditional covariance matrix  process of  $\{  {\bf z}_t \} $ coincides with $\{H_t\} $.

Different prescriptions for the time evolution of the conditional covariance matrix $\{H_t\} $ determine different vector conditional heteroscedastic models. In this paper we will focus on the {\bf VEC-GARCH model} (just VEC in what follows). This model was introduced in~\cite{bollerslev:vec} as the direct generalization of the univariate GARCH model~\cite{bollerslev:garch} in the sense that every conditional variance and covariance is a function of all lagged conditional variances and covariances as well as all squares and cross-products of the lagged time series values. More specifically, the VEC(q,p) model is determined by
\begin{equation*}
\boldsymbol{h}_t={\bf c}+\sum_{i=1}^q A _i\boldsymbol{\eta}_{t-i}+\sum_{i=1}^p B _i \boldsymbol{h}_{t-i},
\end{equation*}
where $\boldsymbol{h}_t:= {\rm vech}(H _t) $, $ \boldsymbol{\eta}_t:=  {\rm vech}({\bf z}_t {\bf z}^T )$, ${\bf c} $ is a $N$-dimensional vector, with $N:=n(n+1)/2 $ and $A _i, B _i \in \mathbb{M}_N $. 

In the rest of the paper we will restrict to the case  $p=q=1$, that is:
\begin{equation}
\label{vec11 model}
\left\{
\begin{array}{ccl}
\mathbf{z}_t &=& H_t^{1/2}\boldsymbol{\epsilon _t}\quad\quad\text{with}\quad\quad
\{\boldsymbol{\epsilon _t}\}\sim {\rm IIDN}({\boldsymbol 0}, {\boldsymbol I}_n),\\
\boldsymbol{h}_t&=&{\bf c}+A \boldsymbol{\eta}_{t-1}+B  \boldsymbol{h}_{t-1}.
\end{array}
\right.
\end{equation}
In this case the model needs $N(2N+1) = \frac{1}{2}(n ^2+ n)(n ^2+n+1)$ parameters for a complete specification. 

\subsection{Positivity and stationarity constraints}

The general prescription for the VEC model spelled out in~(\ref{vec11 model}) does not guarantee that it has stationary solutions. Moreover, as we saw above, the resulting matrices $\{H _t\}_{t \in \mathbb{N}}$ are the conditional covariance matrices of the resulting process and therefore, additional constraints should be imposed on the parameter matrices ${\bf c}, A $,  and $B$  in order to ensure that they are symmetric and positive semidefinite. Unlike the situation encountered in the one-dimensional case, necessary and sufficient conditions for positivity and stationarity seem very difficult to find and we will content ourselves with sufficient specifications.

\medskip

\noindent {\bf Positivity constraints:} we will use the sufficient conditions introduced by Gourieroux in~\cite{gourieroux:book}  that, as we show in the next proposition, can be explicitly formulated using the map $\Sigma $ introduced in Definition~\ref{sigma operator}. 

\begin{proposition}
\label{positivity constraint}
If the parameter matrices ${\bf c}, A$, and $B$ in~(\ref{vec11 model}) are such that ${\rm math}({\bf c}), \Sigma(A) $, and $\Sigma(B) $ are positive semidefinite then so are the resulting   conditional covariance matrices $\{H _t\}_{t \in \mathbb{N}}$, provided the initial condition $H _0 $ is positive semidefinite.
\end{proposition}

\medskip

\noindent {\bf Second order stationarity constraints:} Gourieroux~\cite{gourieroux:book} has stated sufficient conditions in terms of the spectral radius of $A+B $ that we will make more restrictive in order to ensure the availability of a formulation in terms of positive semidefiniteness constraints.

\begin{proposition}
\label{Second order stationarity constraints}
The VEC model specified in~(\ref{vec11 model}) admits a unique second order stationary solution if all the eigenvalues of $A+B $ lie strictly inside the unit circle. This is always the case whenever the top singular eigenvalue $\sigma_{{\rm max}}(A+B)$ of $A+B $ is smaller than one or, equivalently, when the matrix $\mathbb{I}_N-(A+B)(A+B)^T $ is positive definite. If any of these conditions is satisfied, the marginal variance of the model is given by
\begin{equation}
\label{stationary variance}
\Gamma(0)={\rm math}(E[ {\bf h}_t])={\rm math}((\mathbb{I}_N-A-B) ^{-1}{\bf c}).
\end{equation}
\end{proposition}

\subsection{The likelihood function, its gradient, and computability constraints}

Given a sample ${\bf z}=\{{\bf z}_1, \ldots, {\bf z}_T\} $, the quasi-loglikelihood associated to~(\ref{vec11 model}) is:
\begin{equation}
\label{likelihood vec11}
{\rm log}L({\bf z}; \boldsymbol{\theta})=- \frac{TN}{2}\log 2 \pi- \frac{1}{2}\sum_{t=1}^T\log(\det H _t)- \frac{1}{2}\sum_{t=1}^T {\bf z}_t ^TH _t ^{-1}{\bf z} _t
\end{equation}
where $\boldsymbol{\theta}:= \left({\bf c}, A, B\right)$. In this expression, the matrices $H _t $ are constructed out of $\boldsymbol{\theta} $ and the sample ${\bf z} $ using the second expression in~(\ref{vec11 model}). This implies that the dependence of ${\rm log}L $ on $\boldsymbol{\theta} $ takes place through the matrices $H _t$.   Notice that these matrices are well defined once initial values $H _0$ and ${\bf z}_0 $ have been fixed.  This initial values are usually taken out of a presample; if this is not available it is customary to take the mean values associated to the stationary model, namely ${\bf z}= {\bf 0} $ and $H _0={\rm math}((\mathbb{I}_N-A-B) ^{-1}{\bf c}) $  (see~(\ref{stationary variance})). Once the initial conditions have been fixed, it can be shown by induction that
\begin{equation}
\label{explicit formula for h}
{\bf h} _t= \left(\sum_{i=0}^{t-1}B ^i\right){\bf c}+\sum_{i=0}^{t-1}B ^iA \boldsymbol{\eta}_{t-i-1}+B ^t {\bf h} _0.
\end{equation}
The maximum likelihood estimator $\widehat{\boldsymbol{\theta}} $ of $\boldsymbol{\theta}$
is the value that maximizes~(\ref{likelihood vec11}) for a given sample ${\bf z} $. The search of that extremal is carried out using an optimization algorithm that we will discuss later on in the paper and that requires the gradient  $\nabla_ {\boldsymbol{\theta}}{\rm log}L({\bf z}; \boldsymbol{\theta})$ of ${\rm log}L $. In order to compute it  we write the total quasi-loglikelihood as a sum of $T$ conditional loglikelihoods 
\begin{equation*}
l _t({\bf z}_t; A,B,c)=- \frac{N}{2}\log 2 \pi- \frac{1}{2}\log(\det H _t)- \frac{1}{2} {\bf z}_t ^TH _t ^{-1}{\bf z} _t
\end{equation*}A lengthy calculation shows that:
\begin{eqnarray}
\nabla _{{\bf c}}l_t&=&\left[\left(\gamma_t -\Gamma_t \right)^T\sum_{i=0}^{t-1}B ^{i} \right]^T,\label{grad1}\\
\nabla _Al_t&=&\left[\sum_{i=0}^{t-1}\boldsymbol{\eta}_{t-i-1}\left(\gamma_t -\Gamma_t \right)^TB ^i \right]^T,\label{grad2}\\
\nabla _Bl_t&=&\left[\sum_{i=0}^{t-1}\left[\sum_{j=0}^{i-1}B ^j({\bf c}+A\boldsymbol{\eta}_{t-i-1})\left(\gamma_t -\Gamma_t \right)^TB ^{i-j-1}
 +B ^j {\bf h} _0\left(\gamma_t -\Gamma_t \right)^TB ^{t-j-1}\right]\right]^T,\label{grad3}
\end{eqnarray}
where
\begin{equation*}
\Gamma _t:= \frac{1}{2}{\rm math} ^\ast (H _t^{-1}), \quad \gamma _t:= \frac{1}{2}{\rm math} ^\ast (\Lambda _t), \quad\text{and }\quad
\Lambda _t:= H _t^{-1} {\bf z} _t{\bf z} _t^T H _t^{-1}.
\end{equation*}
These formulas for the gradient were obtained by using the explicit expression of the conditional covariance matrices~(\ref{explicit formula for h}) in terms of the sample elements and the coefficient matrices. Such a closed form expression is not always available as soon as the model becomes slightly more complicated; for example, if one adds to the model~(\ref{vec11 model}) a drift term like in~\cite{duan:GARCH:pricing} for the one dimensional GARCH case, an expression like~(\ref{explicit formula for h}) ceases to exist. That is why, in the next proposition, we introduce an alternative iterative method that can be extended to more general models, it is well adapted to its use under the form of a computer code and, more importantly, suggests the introduction of an additional estimation constraint that noticeably shortens the computation time needed for its numerical evaluation.

\begin{proposition}
\label{gradient by recursion}
Let ${\bf z}=\{{\bf z}_1, \ldots, {\bf z}_T\} $ be a sample, $\boldsymbol{\theta}:=({\bf c},A,B) $, and let $ {\rm log}L({\bf z}; \boldsymbol{\theta}) $ be the quasi-loglikelihood introduced in~(\ref{likelihood vec11}). Then, for any component $\theta $ of the three-tuple $\boldsymbol{\theta}$, we have
\begin{equation}
\label{general expression of gradient}
\nabla _{ \boldsymbol{\theta} } \log L=\sum_{t=1}^T \nabla_{\boldsymbol{\theta}} l _t =\sum_{t=1}^T T^\ast _{\boldsymbol{\theta}}   H _t \cdot  \nabla_{H _t} l _t, \quad\text{where}
\end{equation}
\begin{equation}
\label{gradient with respect to Ht}
\nabla_{H _t} l _t= - \frac{1}{2} \left[H _t ^{-1}- H _t ^{-1} {\bf z} _t{\bf z} _t^T H _t ^{-1}  \right],
\end{equation}
and the differential operators $T^\ast _{ \theta } H _t $ are determined by the recursions:
\begin{eqnarray}
T _{{\bf c}}^\ast H _t \cdot \Delta&=& {\rm math}^\ast(\Delta)+T _{{\bf c}}^\ast H_{t-1}\cdot {\rm vech}^\ast(B ^T{\rm math}^\ast(\Delta) ), \label{tht1}\\  
T _A^\ast H _t \cdot \Delta&=& {\rm math}^\ast(\Delta)\cdot \boldsymbol{\eta}_{t-1}^T+T _A^\ast H_{t-1}\cdot {\rm vech}^\ast(B ^T{\rm math}^\ast(\Delta) ),\label{tht2}\\  
T _B^\ast H _t \cdot \Delta&=& {\rm math}^\ast(\Delta)\cdot {\rm vech}(H_{t-1})^T+T _B^\ast H_{t-1}\cdot {\rm vech}^\ast(B ^T{\rm math}^\ast(\Delta) ),\label{tht3}
\end{eqnarray}
with $ \boldsymbol{\eta}_t=  {\rm vech}({\bf z}_t {\bf z}^T )$, $\Delta \in \Bbb S_n$ and setting $T _{{\bf c}}^\ast H _0= {\bf 0}$,  $T _A^\ast H _0=T _B^\ast H _0= {\bf 0} $. The operators $T^\ast _{ \theta } H _t $ constructed in~(\ref{tht1})--(\ref{tht3}) are the adjoints of the partial tangent maps $T _{{\bf c}}H _t: \mathbb{R}^N \rightarrow \Bbb S _n $, $T _{A}H _t: M_N \rightarrow \Bbb S _n $, and $T _{B}H _t: M_N \rightarrow \Bbb S _n $  to $H _t({\bf c},A,B):= {\rm math}({\bf h} _t({\bf c},A,B)) $, with  ${\bf h} _t({\bf c},A,B) $ as defined in~(\ref{explicit formula for h}).
\end{proposition}

\medskip

\noindent {\bf Matrix expression of the recursions~(\ref{tht1})--(\ref{tht3}):} the use of Proposition~\ref{gradient by recursion} requires translating the operator recursions~(\ref{tht1})--(\ref{tht3}) into matrix recursions. In this particular case this can be achieved by writing $\Delta \in \Bbb S _n $  as $\Delta= {\rm vech}^\ast(\mathbf{v})$, with $\mathbf{v} = {\rm math}^\ast( \Delta) \in \mathbb{R}^N$. With this change of variables, the expression~(\ref{tht1}) becomes
\begin{equation}
\label{rec 1}
T _{{\bf c}}^\ast H _t \cdot {\rm vech}^\ast(\mathbf{v})=\mathbf{v}+ T _{{\bf c}}^\ast H _{t-1} \cdot {\rm vech}^\ast(B ^T \mathbf{v}).
\end{equation}
Let $c _t \in \mathbb{M}_N$ be the matrix associated to the linear operator $T _{{\bf c}}^\ast H _t \circ  {\rm vech}^\ast: \mathbb{R}^N \rightarrow \mathbb{R} ^N $. In view of~(\ref{rec 1}), the matrices $\{c _t\}_{t \in \{1, \ldots, T\}}$ are determined by the recursions
\begin{equation}
\label{recuc}
c _t= \mathbb{I} _N+ c_{t-1}B ^T.
\end{equation}
Once the family $\{c _t\}_{t \in \{1, \ldots, T\}}$ has been computed, it can be used in~(\ref{general expression of gradient}) by noticing that 
\begin{equation*}
T _{{\bf c}}^\ast H _t \cdot \Delta=c _t \cdot {\rm math}^\ast (\Delta).
\end{equation*}
Regarding~(\ref{tht2}), let $A _t\in \mathbb{M}_{N ^2,\, N} $ be the matrix associated to the linear operator ${\rm vec} \circ T _{A}^\ast H _t \circ  {\rm vech}^\ast: \mathbb{R}^N \rightarrow \mathbb{R} ^{N ^2} $. Given that 
\begin{equation*}
{\rm vec}(\mathbf{v} \boldsymbol{\eta}^T_{t-1})={\rm vec}(\mathbb{I}_N\mathbf{v} \boldsymbol{\eta}^T_{t-1})= \left(\boldsymbol{\eta}_{t-1}\otimes \mathbb{I}_N\right) {\rm vec}(\mathbf{v})=\left(\boldsymbol{\eta}_{t-1}\otimes \mathbb{I}_N\right) \mathbf{v},
\end{equation*}
the recursion~(\ref{tht2}) implies that the family $\{A _t\}_{t \in \{1, \ldots, T\}}$ is determined by
\begin{equation}
\label{recuA}
A _t= \left(\boldsymbol{\eta}_{t-1}\otimes \mathbb{I}_N\right)+A_{t-1}B ^T
\end{equation}
and hence, once it has been computed, it can be used in~(\ref{general expression of gradient}) by noticing that 
\begin{equation*}
T _{A}^\ast H _t \cdot \Delta={\rm mat} \left(A_t \cdot {\rm math}^\ast (\Delta)\right),
\end{equation*}
where we recall that mat denotes the inverse of the vec operator. Finally, let $B _t\in \mathbb{M}_{N ^2,\, N} $ be the matrix associated to the linear operator ${\rm vec} \circ T _{B}^\ast H _t \circ  {\rm vech}^\ast: \mathbb{R}^N \rightarrow \mathbb{R} ^{N ^2} $. The family $\{B _t\}_{t \in \{1, \ldots, T\}}$ is determined by
\begin{equation}
\label{recuB}
B _t= \left( {\rm vech}(H_{t-1})\otimes \mathbb{I}_N\right)+B_{t-1}B ^T, \quad \mbox{and hence,} \quad T _{B}^\ast H _t \cdot \Delta={\rm mat} \left(B_t \cdot {\rm math}^\ast (\Delta)\right).
\end{equation}

\medskip

\noindent {\bf The computability constraints.} In the particular case of the {\rm VEC}(1,1) model, the existence of the matrix recursions~(\ref{recuc})--(\ref{recuB}) associated to~(\ref{tht1})--(\ref{tht3}) makes the computation of~(\ref{general expression of gradient}) relatively easy. For more general models, the matrix recursions are not easily available and one is forced to work directly with (the analog of)~(\ref{tht1})--(\ref{tht3}). In those cases, if   we deal with a long time series sample, the computation of the gradient~(\ref{general expression of gradient}) may turn out numerically very expensive since it consists of the sum of $T$ terms $T^\ast _{ \theta }  H _t \cdot  \nabla_{H _t} l _t $, each of which is made of the sum of the $t$ terms recursively defined in~(\ref{tht1})--(\ref{tht3}).  A major simplification can be obtained if we restrict ourselves in the estimation process to matrices   $B$ whose top eigenvalue is in norm smaller than one. The defining expressions for the differential operators $T ^\ast _\theta H _t $ show that in that situation,  only a certain number of iterations, potentially small, is needed to compute the gradients with a prescribed precision. This is particularly visible in the expressions~(\ref{grad1})--(\ref{grad3}) where the dependence on the powers of $B$ makes very small many of the involved summands whenever the spectrum of $B$ is strictly contained in the unit disk. This is the reason why we will impose this as an additional estimation constraint. The details of this statement are spelled out in the proposition below that we present after the summary of the constraints that we will impose all along the paper on the model~(\ref{vec11 model}):
\begin{description}
\item [(SC)] {\bf Stationarity constraints:} $\mathbb{I}_N(1- \epsilon_{AB})-(A+B)(A+B)^T\succeq 0 $ for some small $\epsilon_{AB}>0 $.
\item [(PC)] {\bf Positivity constraints:} ${\rm math}({\bf c})- \epsilon_{{\bf c}}\mathbb{I} _n\succeq 0 $, $\Sigma(A)- \epsilon_A \mathbb{I}_{n ^2}\succeq 0 $, and $\Sigma(B)- \epsilon_B \mathbb{I}_{n ^2}\succeq 0  $, for some small $\epsilon_A, \epsilon_B, \epsilon_{{\bf c}} >0 $.
\item [(CC)] {\bf Computability constraints:} $\mathbb{I}_N(1- \widetilde{\epsilon}_{B})-BB^T\succeq 0 $ for some small $\widetilde{\epsilon}_{B}>0 $.
\end{description}
 
\begin{proposition}
\label{estimate number iterations}
Let $t \in \mathbb{N} $  be a fixed lag and let $T^\ast _{ \theta } H _t $ be the differential operators defined by applying $t$ times the recursions~(\ref{tht1})-(\ref{tht3}). Consider now the operators $T^\ast _{ \theta } H _t ^k$ obtained by truncating the  recursions~(\ref{tht1})-(\ref{tht3}) after $k$ iterations, $k< t $. If we assume that the coefficients  ${\bf c}, A $, and $B$ satisfy the constraints {\bf (SC)}, {\bf (PC)}, and {\bf (CC)} then the error committed in the truncations can be estimated using the following inequalities satisfied by the operator norms:
\begin{eqnarray}
\|T^\ast _{{\bf c} } H _t -T^\ast _{ {\bf c} } H _t ^k\|_{{\rm op}}&\leq & \frac{2(1- \widetilde{\epsilon } _B)^k}{\widetilde{\epsilon } _B},\label{ub1}\\
\|E\left[T^\ast _{A } H _t -T^\ast _{ A } H _t ^k\right]\|_{{\rm op}}&\leq & \frac{2(1- \widetilde{\epsilon } _B)^k\| {\bf c}\|}{\epsilon _{AB}},\label{ub2}\\
\|E\left[T^\ast _{B} H _t -T^\ast _{B} H _t ^k\right]\|_{{\rm op}}&\leq & \frac{2(1- \widetilde{\epsilon } _B)^k\| {\bf c}\|}{\epsilon _{AB}}.\label{ub3}
\end{eqnarray}
Notice that the last two inequalities estimate the error committed in mean. As consequence of these relations, if we allow a maximum expected error $\delta$ in the computation of the gradient~(\ref{general expression of gradient}) then a lower bound for the number $k$ of iterations that need to be carried out in~(\ref{tht1})--(\ref{tht3}) is:
\begin{equation}
\label{k estimate}
k=\max \left\{\frac{\log \left(\frac{\widetilde{\epsilon} _B\delta}{2}\right)}{\log (1- \widetilde{\epsilon} _B)}, \frac{\log \left(\frac{\widetilde{\epsilon} _B\epsilon_{AB}\delta}{2 \epsilon_{{\bf c}}}\right)}{\log (1- \widetilde{\epsilon} _B)}\right\}.
\end{equation}
\end{proposition}

\begin{remark}
\normalfont
The estimate~(\ref{k estimate}) for the minimum number of iterations needed to reach a certain precision in the computation of the gradient is by no means sharp. Numerical experiments show that the figure produced by this formula is in general too conservative. Nevertheless, this expression is still very valuable for it explicitly shows the pertinence of the computability constraint {\bf (CC)}.
\end{remark}

\begin{remark}
\normalfont
We emphasize that the constraints {\bf (SC)}, {\bf (PC)}, and {\bf (CC)} are sufficient conditions for stationarity, positivity, and computability, respectively, but by no means necessary. For example {\bf (SC)} and {\bf (CC)} could be replaced by the more economical (but also more restrictive) condition that imposes $A, B \in \Bbb S_N^+ $ with $\lambda_{{\rm max}}(A+B)\leq(1- \epsilon_{AB})$. In this situation it can be easily shown that $\lambda_{{\rm max}}(B)<1 $ and hence the computability constrained is automatically satisfied.
\end{remark}

\section{Calibration via Bregman matrix divergences}
\label{Calibration via Bregman matrix divergences}

In this section we present an efficient optimization method that, given a sample ${\bf z}$, provides the parameter value $\widehat{\boldsymbol{\theta}} $ corresponding to the {\rm VEC}(1,1) model that fits it best by maximizing the quasi-loglikelihood~(\ref{likelihood vec11}) subjected to the constraints {\bf (SC)}, {\bf (PC)}, and {\bf (CC)}. It can be proved under certain regularity hypotheses (see~\cite[page 119]{gourieroux:book}) that the quasi-loglikelihood estimator $\widehat{\boldsymbol{\theta}} $ is consistent and asymptotically normal:
\begin{equation}
\label{estimator properties 1}
\sqrt{T}(\widehat{\boldsymbol{\theta}}- \boldsymbol{\theta}_0)\xrightarrow[{\rm dist}]{}N(0, \Omega_0) \quad \mbox{where} \quad \Omega_0= A _0^{-1}B _0A _0^{-1}, \quad \mbox{with} \quad
\end{equation}
\begin{equation}
\label{estimator properties 2}
A _0= E_{\boldsymbol{\theta}_0} \left[- \frac{\partial ^2 l _t (\boldsymbol{\theta}_0)}{\partial \boldsymbol{\theta}\partial \boldsymbol{\theta}^T}\right]\quad \mbox{ and } \quad B _0= E_{\boldsymbol{\theta}_0} \left[\frac{\partial l _t (\boldsymbol{\theta}_0)}{\partial \boldsymbol{\theta}}\frac{\partial l _t (\boldsymbol{\theta}_0)}{\partial \boldsymbol{\theta}^T}\right].
\end{equation}
These matrices are usually consistently estimated by replacing the expectations by their empirical means and the true value of the parameter $\boldsymbol{\theta}_0 $ by the estimator $ \widehat{\boldsymbol{\theta}}$:
\begin{equation*}
\widehat{A} _0= - \frac{1}{T}\sum_{i=1}^T\frac{\partial ^2 l _t (\widehat{\boldsymbol{\theta}})}{\partial \boldsymbol{\theta}\partial \boldsymbol{\theta}^T}, \qquad \widehat{B} _0= \frac{1}{T} \sum_{i=1}^T\frac{\partial l _t (\widehat{\boldsymbol{\theta}})}{\partial \boldsymbol{\theta}}\frac{\partial l _t (\widehat{\boldsymbol{\theta}})}{\partial \boldsymbol{\theta}^T}.
\end{equation*}

\subsection{Constrained optimization via Bregman divergences}
\label{Constrained optimization via Bregman divergences}

The optimization method that we will be carrying out to maximize the quasi-loglikelihood is based on the use of {\bf Burg's matrix divergence}. This divergence is  presented, for example, in~\cite{kulis:sustik:dhillon} and it is a particular instance of a Bregman divergence. Bregman divergences are of much use in the context of machine learning (see for instance~\cite{dhillon:tropp, kulis:sra:dhillon} and references therein). 

In our situation we have opted for this technique as it allows for a particularly efficient treatment of the constraints in our problem, avoiding the need to solve additional secondary optimization problems. In order to make this more explicit it is worth mentioning that we also considered different approaches consisting of optimizing quadratically penalized 
local first or second order models with the positive 
semidefinite constraints {\bf (PS)}, {\bf  (SC)},  and {\bf (CC)}; since we were not able to find a closed form expression for the optimization step induced by this constrained local model, 
we were forced to use Lagrange duality. Even though the constraints admit a simple conic formulation that allowed us to explicitly formulate the problem, this approach finally resulted in a problem that is much more computationally demanding  than
just incorporating the constraints into the primal scheme using Bregman divergences, as we propose below.

\begin{definition}
Let $X,Y \in \Bbb S_n $ and $\phi: \Bbb S_n\longrightarrow \mathbb{R} $ a strictly convex differentiable function. The {\bf Bregman matrix divergence} associated to $\phi$  is defined by
\begin{equation*}
D_\phi(X,Y):= \phi(X)- \phi(Y)- \operatorname{trace} \left(\nabla \phi(Y) ^T(X-Y) \right).
\end{equation*}
\end{definition}
Bregman divergences are used to measure distance between matrices. Indeed, if we take the squared Frobenius norm as the function $\phi $, that is $\phi(X):=\| X\| ^2  $, then $D_\phi(X,Y):=\| X-Y\| ^2  $. Other example is the {\bf von Neumann divergence} which is the Bregman divergence associated to the entropy of the eigenvalues of a positive definite matrix; more explicitly, if $X$ is a positive definite matrix with eigenvalues $\{ \lambda _1, \ldots, \lambda _n\} $, then $\phi(X):=\sum_{i=1}^n (\lambda _i\log \lambda _i- \lambda_i) $. In our optimization problem we will be using {\bf Burg's matrix divergence} (also called the {\bf LogDet divergence} or {\bf Stein's loss} in the statistics literature~\cite{james:stein}) which is the Bregman divergence obtained out of the Burg entropy of the eigenvalues of a positive definite matrix, that is $\phi(X):=-\sum_{i=1}^n\log \lambda _i $, or equivalently  $\phi(X):= -\log\det (X) $. The resulting Bregman divergence over positive definite matrices is 
\begin{equation}
\label{burg divergence}
D_B(X,Y):= \operatorname{trace}(XY ^{-1})-\log\det(XY ^{-1})-n.
\end{equation}
The three divergences that we just introduced are examples of {\bf spectral} divergences, that is, the function $\phi $ that defines them can be written down as the composition $\phi=\varphi \circ \lambda $, where $ \varphi: \mathbb{R}^n\longrightarrow \mathbb{R}  $ is differentiable strictly convex function and $\lambda :\Bbb S_n \longrightarrow  \mathbb{R}^n $ is the function that  lists the eigenvalues of $X$ in algebraically decreasing order. It can be seen (see Appendix A in~\cite{kulis:sustik:dhillon}) that spectral Bregman matrix divergences are invariant by orthogonal conjugations, that is, for any orthogonal matrix $Q \in \mathbb{O}_n $:
\begin{equation*}
D_{\phi}(Q ^TXQ,Q ^TYQ)=D_{\phi}(X,Y).
\end{equation*}
Burg divergences are invariant by an even larger group since
\begin{equation*}
D_{B}(M ^TXM,M ^TYM)=D_{B}(X,Y),
\end{equation*}
for any square non-singular matrix $M $. Additionally, for any non-zero scalar $\alpha$:  
\begin{equation*}
D_{B}(\alpha X,\alpha Y)=D_{B}(X,Y).
\end{equation*}

\begin{figure}[htp]
\includegraphics[scale=.33]{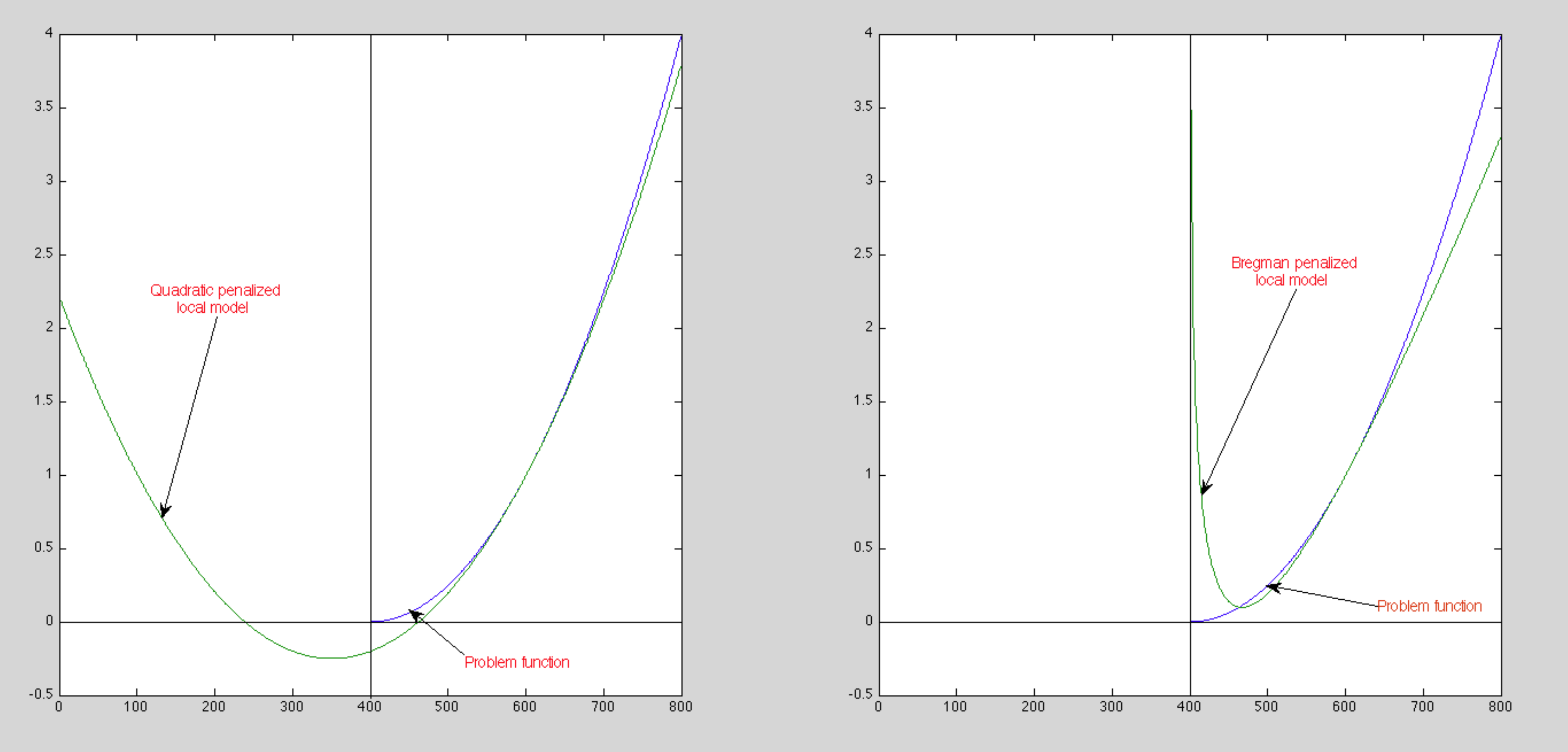}
\caption{The blue function is subjected to the constraint $x\geq 400$ and, being strictly increasing, attains its minimum at $x=400 $. On the left hand side we use a standard quadratically penalized local model of the function and we see that its minimum is attained outside the constrained domain. On the right hand side we replace the quadratic penalization by a Bregman divergence that forces the local model to exhibit its optimum at a point that satisfies the constraints.}
\label{fig:localmodel}
\end{figure}

The use of Bregman divergences in matrix constrained optimization problems is substantiated by replacing the quadratic term in the local model,  that generally uses the Frobenius distance, by a Bregman divergence that places the set outside the constraints at an infinite distance. More explicitly, suppose that the constraints of a optimization problem are formulated as a positive definiteness condition $A\succeq 0 $ and that we want to find
\begin{equation*}
\mathop{\rm arg\, min}_{A\succeq 0}\, f(A),
\end{equation*}
by iteratively solving the optimization problems associated to penalized local models of the form 
\begin{equation}
\label{local model}
f_{A^{(n)}}(A):=f \left(A^{(n)}\right)+\left\langle \nabla f\left(A^{(n)}\right),A-A^{(n)}\right\rangle+ \frac{L}{2} D_\phi(A,A^{(n)}). 
\end{equation}
If in this local model we take $\phi(X):=\| X\| ^2  $ and the elastic penalization constant $L$ is small enough, the minimizer $\mathop{\rm arg\, min}_{A\succeq 0}\, f_{A^{(n)}}(A) $ is likely to take place outside the constraints. However, if we use Burg's divergence $D_B $  instead, and $A ^{(n)}$ is positive definite, then so is  $\mathop{\rm arg\, min}_{A\succeq 0}\, f_{A^{(n)}}(A) $ for no matter what value of the parameter $L$. This is so because as $A $ approaches the constraints, the term $D_\phi(A,A^{(n)}) $ becomes increasingly close to infinity producing the effect that we just described; see Figure~\ref{fig:localmodel} for an illustration. The end result of using Bregman divergences is that {\it they reduce a constrained optimization problem to a series of local unconstrained ones}.

\subsection{Bregman divergences for VEC models}
\label{Bregman divergences for VEC models}

Before we tackle the VEC  estimation problem, we add to  {\bf (SC)}, {\bf (PC)}, and {\bf (CC)} a fourth constraint on the variable ${\bf c}  \in \mathbb{R}^N$ that makes compact the optimization domain:
\begin{description}
\item [(KC)] {\bf Compactness constraint:} $K \mathbb{I}_N- {\rm math}({\bf c})\succeq 0  $ for some $K \in \mathbb{R}$.
\end{description}
In practice the constant $K$ is taken as a multiple of the Frobenius norm of the covariance matrix of the sample. This is a reasonable choice since by~(\ref{stationary variance}), in the stationary regime ${\bf c}= (\mathbb{I}_N-A-B){\rm vech}(\Gamma (0))$; moreover, by the constraint ${\bf (SC)} $ and~(\ref{operator norm vech}) we have
\begin{equation*}
\|{\bf c}\|= \|(\mathbb{I}_N-A-B){\rm vech}(\Gamma (0))\|\leq \|\mathbb{I}_N-A-B\|_{{\rm op}}\| {\rm vec}\|_{{\rm op}}\|\Gamma (0)\|\leq 2 \|\Gamma (0)\|.
\end{equation*}

Now, given a sample ${\bf z} $  and a starting value for the parameters $ \boldsymbol{\theta}_0=({\bf c}_0, A _0 , B _0)$, our goal is  finding the minimizer of minus the quasi-loglikelihood $f(\boldsymbol{\theta}):=-{\rm log}L({\bf z}; \boldsymbol{\theta}) $, subjected to the constraints {\bf (SC)}, {\bf (PC)}, {\bf (CC)}, and {\bf (KC)}. We will worry about the problem of finding a preliminary estimation $\boldsymbol{\theta} _0 $ later on in Section~\ref{Preliminary estimation}. As we said before, our method is based on recursively optimizing penalized local models that incorporate Bregman divergences that ensure that the constraints are satisfied. More specifically, the estimate of the optimum $\boldsymbol{\theta}^{(n+1)} $ after $n$ iterations is obtained by solving
\begin{equation}
\label{local optimization problem}
\boldsymbol{\theta}^{(n+1)} = \mathop{\rm arg\, min}_{\boldsymbol{\theta}\in \mathbb{R}^N\times \mathbb{M}_N \times \mathbb{M}_N ,}\, \tilde{f}^{(n)}(\boldsymbol{\theta}),
\end{equation}
where $\tilde{f}^{(n)} $ is defined by:
\begin{eqnarray}
\tilde{f}^{(n)}(\boldsymbol{\theta}) &= &f(\boldsymbol{\theta}^{(n)})+\langle\nabla f(\boldsymbol{\theta}^{(n)} ), \boldsymbol{\theta}-\boldsymbol{\theta}^{(n)} \rangle+ \frac{L _1}{2} D_B(\mathbb{I}_N-(A+B)^T(A+B),\mathbb{I}_N-(A^{(n)}+B^{(n)})^T(A^{(n)}+B^{(n)}))\notag\\
	&+ & \frac{L _2}{2} D_B(\Sigma(A), \Sigma(A^{(n)}))+\frac{L _3}{2} D_B(\Sigma(B), \Sigma(B^{(n)}))+\frac{L _4}{2} D_B(\mathbb{I}_N-B^TB,\mathbb{I}_N-B^{(n)\,T}B^{(n)})\notag\\
	&+&\frac{L _5}{2} D_B({\rm math}({\bf c}), {\rm math}({\bf c}^{(n)}))+\frac{L _6}{2} D_B(K \mathbb{I}_N-{\rm math}({\bf c}), K \mathbb{I}_N-{\rm math}({\bf c}^{(n)})).\label{definition local model}
\end{eqnarray}
Notice that for the sake of simplicity we have incorporated the constraints in the divergences with the constraint tolerances $\epsilon_{AB}, \epsilon_A, \epsilon _B, \widetilde{\epsilon}_B$, and $\epsilon_{{\bf c}}$ set equal to zero.

The local optimization problem in~(\ref{local optimization problem}) is solved by finding the value $\boldsymbol{\theta}_0 $ for which  
\begin{equation}
\label{solution local model}
\nabla \tilde{f}^{(n)}(\boldsymbol{\theta}_0) =0. 
\end{equation}
A long but straightforward computation shows that the gradient $\nabla \tilde{f}^{(n)}(\boldsymbol{\theta}) $ is given by the expressions:
\begin{eqnarray}
\nabla _A\tilde{f}^{(n)}(\boldsymbol{\theta})&= &\nabla_A f(\boldsymbol{\theta}^{(n)} )-L _1(A+B) \left(\left(\mathbb{I}_N-(A^{(n)}+B^{(n)})^T(A^{(n)}+B^{(n)}) \right)^{-1}- \left( \mathbb{I}_N-(A+B)^T(A+B)\right)^{-1}\right)\notag\\
	&  & +\frac{L _2}{2} \Sigma^\ast  \left(\Sigma(A^{(n)})^{-1}-\Sigma(A)^{-1}\right),\label{gradi1}\\
\nabla _B\tilde{f}^{(n)}(\boldsymbol{\theta})&= &\nabla_B f(\boldsymbol{\theta}^{(n)} )-L _1(A+B) \left(\left(\mathbb{I}_N-(A^{(n)}+B^{(n)})^T(A^{(n)}+B^{(n)}) \right)^{-1}- \left( \mathbb{I}_N-(A+B)^T(A+B)\right)^{-1}\right)\notag\\
	&  & +\frac{L _3}{2} \Sigma^\ast  \left(\Sigma(B^{(n)})^{-1}-\Sigma(B)^{-1}\right)-L _4B \left(\left(\mathbb{I}_N-B^{(n)\, T}B^{(n)} \right)^{-1}- \left( \mathbb{I}_N-B^TB\right)^{-1}\right),\label{gradi2}\\
\nabla _{{\bf c}}	\tilde{f}^{(n)}(\boldsymbol{\theta})&= &\nabla _{{\bf c}} f(\boldsymbol{\theta}^{(n)} )+ \frac{L _5}{2} {\rm math}^\ast\left({\rm math}({\bf c}^{(n)})^{-1}-{\rm math}({\bf c})^{-1}\right)\notag\\
	& &- \frac{L _6}{2}{\rm math}^\ast\left((K \mathbb{I}_N-{\rm math}({\bf c}^{(n)}))^{-1}-(K \mathbb{I}_N-{\rm math}({\bf c}))^{-1}\right),\label{gradi3}
\end{eqnarray} 
where $\nabla_{\boldsymbol{\theta}} f(\boldsymbol{\theta}^{(n)} ) =-\nabla_{\boldsymbol{\theta}} {\rm log}L({\bf z}; \boldsymbol{\theta}^{(n)})$ is provided by the expressions in Proposition~\ref{gradient by recursion}. We will numerically find the solution of the equation~(\ref{solution local model}) using the Newton-Raphson algorithm, which requires computing the tangent map to $\nabla \tilde{f}^{(n)}(\boldsymbol{\theta}) $. In order to do so, let $g _1 ^{(n)}(A,B)$, $g _2 ^{(n)}(A,B) $, and $g _3 ^{(n)}({\bf c})$  be the functions in the right hand side of the expressions~(\ref{gradi1}),~(\ref{gradi2}), and (\ref{gradi3}), respectively, and $g ^{(n)}(A,B, {\bf c}):= \left(g _1 ^{(n)}(A,B),g _2 ^{(n)}(A,B), g _3 ^{(n)}({\bf c}) \right)$; additionally, define the map $\Lambda(A): \mathbb{M}_N \rightarrow \mathbb{M} _N $ by $\Lambda(A):= \mathbb{I} _N-A ^T A $, as well as 
\begin{eqnarray}
\Xi_{A}^{(n)}(\Delta)&= &- \Delta\left(\Lambda \left(A ^{(n)}\right)^{-1}- \Lambda(A)^{-1}\right)+A\Lambda (A)^{-1}\left(\Delta^T(A)+A^T \Delta\right)\Lambda(A)^{-1},\label{xi function}\\
\mathfrak{X}_A(\Delta)&=&\Sigma^\ast  \left(\Sigma(A)^{-1} \Sigma(\Delta)\Sigma(A)^{-1}\right),\label{german x formula}
\end{eqnarray}
for any $A, \Delta \in \mathbb{M} _N  $. A straightforward computation shows that:
\begin{eqnarray}
T_{(A,B)}g _1^{(n)}\cdot (\Delta_A, \Delta_B) &= &\left(L _1\Xi_{A+B}^{(n)}(\Delta_A)+ \frac{L _2}{2}\mathfrak{X}_A(\Delta_A), L _1\Xi_{A+B}^{(n)}(\Delta_B)\right), \label{jacobian1}\\
T_{(A,B)}g _2^{(n)}\cdot (\Delta_A, \Delta_B) &= &\left(L _1\Xi_{A+B}^{(n)}(\Delta_A),L _1\Xi_{A+B}^{(n)}(\Delta_B)+ \frac{L _3}{2}\mathfrak{X}_B(\Delta_B)+L _4\Xi_{B}^{(n)}(\Delta_B)\right), \label{jacobian2}\\
T_{{\bf c}}g _3^{(n)} \cdot \Delta_{{\bf c}}&= &\frac{L _5}{2} {\rm math}^\ast\left({\rm math}({\bf c})^{-1}{\rm math}(\Delta_{{\bf c}}){\rm math}({\bf c})^{-1}\right)\notag\\
& &+\frac{L _6}{2}{\rm math}^\ast\left((K \mathbb{I}_N-{\rm math}({\bf c}))^{-1}{\rm math}(\Delta_{{\bf c}})(K \mathbb{I}_N-{\rm math}({\bf c}))^{-1}\right).\label{jacobian3}
\end{eqnarray}
In obtaining these equalities we used that the tangent map to the matrix inversion operation ${\rm inv} (X):= X ^{-1} $ is given by $T_X {\rm inv}  \cdot \Delta=- X ^{-1}\Delta X ^{-1} $ and hence
\begin{equation*}
T _A \Lambda(A)^{-1} \cdot \Delta_A= \Lambda(A)^{-1}\left(\Delta_A^T A+A ^T\Delta_A\right)\Lambda(A)^{-1}\quad \mbox{and} \quad T _A\Sigma(A)^{-1}\cdot \Delta_A=- \Sigma(A)^{-1}\Sigma(\Delta_A)\Sigma(A)^{-1}.
\end{equation*}
The use of the tangent maps~(\ref{jacobian1})--(\ref{jacobian3}) in a numerical routine that implements the Newton-Raphson method requires computing the matrix associated to the linear map $T_{(A,B, {\bf c})}g ^{(n)} $. A major part in this task, namely the matrix associated to the map $\Xi ^{(n)} $ in~(\ref{xi function}), admits a closed form expression that avoids a componentwise computation. Indeed:
\begin{eqnarray}
{\rm vec}(\Xi^{(n)}_A(\Delta)) &= & -\left[\left(\Lambda \left(A ^{(n)}\right)^{-1}- \Lambda(A)^{-1}\right)\otimes \mathbb{I}_N\right]{\rm vec}(\Delta)+ \left[\left(A \Lambda(A)^{-1}\right)^T\otimes A \Lambda(A)^{-1}\right]K_{NN}{\rm vec}(\Delta)\notag\\
	& &+\left[\Lambda(A)^{-1\, T}\otimes A \Lambda(A)^{-1} A ^T\right]{\rm vec}(\Delta),\label{for matrix}
\end{eqnarray}
where $K_{NN}$  is the  $(N,N)$-commutation matrix (see~\cite{magnus:neudecker}). This expression implies that the matrix $\widetilde{\Xi^{(n)}_A} \in \mathbb{M} _{N ^2}$ associated to the linear map $\Xi^{(n)}_A: \mathbb{M}_N \rightarrow \mathbb{M}_N  $ is given by
\begin{equation}
\label{part matrix of xi}
\widetilde{\Xi^{(n)}_A}=-\left[\left(\Lambda \left(A ^{(n)}\right)^{-1}- \Lambda(A)^{-1}\right)\otimes \mathbb{I}_N\right]+ \left[\left(A \Lambda(A)^{-1}\right)^T\otimes A \Lambda(A)^{-1}\right]K_{NN}+\left[\Lambda(A)^{-1\, T}\otimes A \Lambda(A)^{-1} A ^T\right].
\end{equation}
In order to obtain~(\ref{for matrix}),  we used the following properties of the {\rm vec} operator:
\begin{equation*}
{\rm vec}(AB)= \left(B ^T\otimes \mathbb{I}\right) {\rm vec}(A), \quad {\rm vec}(ABC)= \left(C ^T\otimes A\right) {\rm vec}(A)\quad \mbox{and} \quad {\rm vec}(A ^T)=K_{NN} {\rm vec}(A),
\end{equation*}
for any $A,B,C \in \mathbb{M}_N$. We have not found a closed formula for the matrices associated to the other linear maps that constitute~(\ref{jacobian1})--(\ref{jacobian3}) and hence they need to be obtained in a componentwise manner by applying them to all the elements of a canonical basis. Let $\widetilde{\mathfrak{X} _A} $, $\widetilde{T_{(A,B, {\bf c})} g ^{(n)}}$, and $\widetilde{T_{{\bf c}} g_3 ^{(n)}}$ be the matrices associated to $\mathfrak{X} _A$, $T_{(A,B, {\bf c})} g ^{(n)}$, and $T_{{\bf c}} g_3 ^{(n)}$, respectively. Then, by~(\ref{jacobian1})--(\ref{jacobian3}), we have:
\begin{equation}
\label{matrix of jacobian}
\widetilde{T_{(A,B, {\bf c})} g ^{(n)}}=
\left(
\begin{array}{ccc}
L _1\widetilde{\Xi^{(n)}_{A+B}}+ \frac{L _2}{2}\widetilde{\mathfrak{X} _A} &L _1\widetilde{\Xi^{(n)}_{A+B}} &\boldsymbol{0}\\
L _1\widetilde{\Xi^{(n)}_{A+B}} &L _1\widetilde{\Xi^{(n)}_{A+B}} + \frac{L _3}{2}\widetilde{\mathfrak{X} _B}+L _4 \widetilde{\Xi^{(n)}_{B}}&\boldsymbol{0}\\
\boldsymbol{0} &\boldsymbol{0} &\widetilde{T_{{\bf c}} g_3 ^{(n)}}
\end{array}
\right).
\end{equation}
Using this matrix, the solution $\boldsymbol{\theta} _0 $ of~(\ref{solution local model}) is the limit of the sequence $\{\boldsymbol{\theta}^{(n,\,k)}\}_{k \in \mathbb{N}}:=\{(A ^{(n,\,k)},B ^{(n,\,k)}, {\bf c}^{(n,\,k)})\}_{k \in \mathbb{N}}$ constructed using the prescription $\boldsymbol{\theta}^{(n,\,1)}=\boldsymbol{\theta}^{(n)}=(A ^{(n)},B ^{(n)}, {\bf c}^{(n)}) $ and by iteratively solving the linear systems:
\begin{equation}
\label{newton raphson iteration}
\widetilde{T_{\boldsymbol{\theta}^{(n,\,k)}} g ^{(n)}}\cdot 
\left(
\begin{array}{c}
{\rm vec}(A^{(n,\,k+1)})\\
{\rm vec}(B^{(n,\,k+1)})\\
{\bf c}^{(n,\,k+1)}
\end{array}
\right)=
-{\rm vec}\left(\nabla \tilde{f}^{(n)}(\boldsymbol{\theta}^{(n,\,k)})\right)+
\widetilde{T_{\boldsymbol{\theta}^{(n,\,k)}} g ^{(n)}}\cdot 
\left(
\begin{array}{c}
{\rm vec}(A^{(n,\,k)})\\
{\rm vec}(B^{(n,\,k)})\\
{\bf c}^{(n,\,k)}
\end{array}
\right).
\end{equation}
\begin{remark}
\normalfont
Since  the tangent map $T_{\boldsymbol{\theta}}g ^{(n)} $ can be assimilated to  the Hessian of $\tilde{f}^{(n)} $, its matricial expression $\widetilde{T_{\boldsymbol{\theta}}g ^{(n)} }$ in~(\ref{matrix of jacobian}) should be symmetric. When this matrix is actually numerically constructed, the part resulting from the matrix identity~(\ref{part matrix of xi}) is automatically symmetric. The rest, that comes out of a componentwise study, may introduce numerical differences that slightly spoil symmetricity and that, in practice, has a negative effect in the performance of the optimization algorithm as a whole. That is why we strongly advice to symmetrize by hand $\widetilde{T_{\boldsymbol{\theta}}g ^{(n)} }$ once it has been computed. \end{remark}

\subsection{Performance improvement: BFGS and trust-region corrections}
\label{Performance improvement: BFGS and trust-region corrections}

The speed of convergence of the estimation algorithm presented in the previous section can be significantly increased by enriching the local model with a quadratic BFGS (Broyden-Fletcher-Goldfarb-Shanno) type term and by only accepting steps of a certain quality  measured by the ratio between the actual descent and that predicted by the local model (see~\cite{trust:region} and references therein).

The BFGS correction is introduced by adding to the local penalized model $\tilde{f}^{(n)} (\boldsymbol{\theta})$ defined in~(\ref{definition local model}), the BFGS Hessian proxy $H ^{(n)} $ iteratively defined by:
\begin{equation*}
H^{(n)}=H ^{(n-1)}+ \frac{y^{(n-1)}y^{(n-1)\, T}}{y^{(n-1)\, T}s^{(n-1)}}- \frac{H ^{(n-1)}s ^{(n-1)}s ^{(n-1)\, T}H ^{(n-1)}}{s ^{(n-1)\, T}H ^{(n-1)}s ^{(n-1)}}.
\end{equation*}
with $H^{(0)} $ an arbitrary positive semidefinite matrix and where $s ^{(n-1)}:= \boldsymbol{\theta}^{(n)}- \boldsymbol{\theta}^{(n-1)} $ and $y ^{(n-1)}:=\nabla f (\boldsymbol{\theta}^{(n)})-\nabla f (\boldsymbol{\theta}^{(n-1)}) $.
More specifically, we replace the local penalized model $\tilde{f}^{(n)} (\boldsymbol{\theta})$ by
\begin{equation*}
\hat{f}^{(n)} (\boldsymbol{\theta}):=\tilde{f}^{(n)} (\boldsymbol{\theta})+ \frac{1}{2}\left(\boldsymbol{\theta}-\boldsymbol{\theta}^{(n)}\right)^TH^{(n)}\left(\boldsymbol{\theta}-\boldsymbol{\theta}^{(n)}\right),
\end{equation*}
whose gradient is obviously given by:
\begin{equation*}
\hat{g}^{(n)}(\boldsymbol{\theta}):=\nabla \hat{f}^{(n)} (\boldsymbol{\theta})=\nabla\tilde{f}^{(n)} (\boldsymbol{\theta})+H^{(n)}\left(\boldsymbol{\theta}-\boldsymbol{\theta}^{(n)}\right)=\widetilde{g}^{(n)} (\boldsymbol{\theta})+H^{(n)}\left(\boldsymbol{\theta}-\boldsymbol{\theta}^{(n)}\right),
\end{equation*}
with $\widetilde{g}^{(n)} (\boldsymbol{\theta})=\nabla\tilde{f}^{(n)} (\boldsymbol{\theta})$ given by~(\ref{gradi1})--(\ref{gradi3}). Using this corrected local penalized model, the solution of the optimization problem will be obtained by iteratively computing
\begin{equation}
\label{local optimization problem bfgs}
\boldsymbol{\theta}^{(n+1)} = \mathop{\rm arg\, min}_{\boldsymbol{\theta}\in \mathbb{R}^N\times \mathbb{M}_N \times \mathbb{M}_N ,}\, \hat{f}^{(n)}(\boldsymbol{\theta}).
\end{equation}
This is carried out by finding the solution $\boldsymbol{\theta}_0 $ of the equation
\begin{equation}
\label{solution local model bfgs}
\hat{g}^{(n)}(\boldsymbol{\theta}_0)=\widetilde{g}^{(n)} (\boldsymbol{\theta}_0)+H^{(n)}\left(\boldsymbol{\theta}_0-\boldsymbol{\theta}^{(n)}\right) =0. 
\end{equation}
using a modified version of the Newton-Raphson iterative scheme spelled out in~(\ref{newton raphson iteration}). Indeed, it is easy to show that $\boldsymbol{\theta}_0 $ is the limit of the sequence $\{\boldsymbol{\theta}^{(n,\,k)}\}_{k \in \mathbb{N}}$ constructed exactly as in Section~\ref{Bregman divergences for VEC models} where the linear systems~(\ref{newton raphson iteration}) are replaced by
\begin{equation}
\label{newton raphson iteration bfgs}
\left(\widetilde{T_{\boldsymbol{\theta}^{(n,\,k)}} g ^{(n)}}+\widetilde{H ^{(n)}}\right)\cdot \widetilde{\boldsymbol{\theta} ^{(n,\,k+1)}}
=
-{\rm vec}\left(\nabla \tilde{f}^{(n)}(\boldsymbol{\theta}^{(n,\,k)})\right)+\widetilde{H ^{(n)}}\cdot \widetilde{\boldsymbol{\theta} ^{(n)}}+
\widetilde{T_{\boldsymbol{\theta}^{(n,\,k)}} g ^{(n)}}\cdot \widetilde{\boldsymbol{\theta} ^{(n,\,k)}}.
\end{equation}
where $\widetilde{\boldsymbol{\theta} ^{(n,\,k+1)}}=\left(
\begin{array}{c}
{\rm vec}(A^{(n,\,k+1)})\\
{\rm vec}(B^{(n,\,k+1)})\\
{\bf c}^{(n,\,k+1)}
\end{array}
\right) $ and $\widetilde{H ^{(n)}} \in \mathbb{M}_{2N ^2+N}$ denotes the matrix associated to $H ^{(n)} $ that satisfies
\begin{equation*}
{\rm vec}\left( H ^{(n)}\cdot \boldsymbol{\theta} \right)=\widetilde{H ^{(n)}} \cdot 
\left( 
\begin{array}{c}
{\rm vec}(A )\\
{\rm vec}(B )\\
{\bf c} 
\end{array}
\right)\quad \mbox{for any} \quad \boldsymbol{\theta}=(A,B, {\bf c}). 
\end{equation*}

\medskip

\noindent {\bf Important remark: the Newton-Raphson method and the constraints.} In Section~\ref{Constrained optimization via Bregman divergences} we explained how the use of Bregman divergences ensures that at each iteration, the extremum of the local penalized model satisfies the constraints of the problem. However, the implementation of the Newton-Raphson method that provides the root of the equation~(\ref{solution local model bfgs}) does not, in general, respect the constraints, and hence this point requires special care. 

In the construction of our optimization algorithm we have used the following prescription in order to ensure that all the elements of the sequence $\{\boldsymbol{\theta}^{(n,\,k)}\}_{k \in \mathbb{N}}$ that converge to the root $\boldsymbol{\theta}_0 $ satisfy the constraints: given $\boldsymbol{\theta}^{(n,\,1)}=\boldsymbol{\theta}^{(n)} $ (that satisfies the constraints) let $\boldsymbol{\theta}^{(n,\,2)} $ be the second value in the Newton-Raphson sequence obtained by solving the linear system~(\ref{newton raphson iteration bfgs}). If the value $\boldsymbol{\theta}^{(n,\,2)} $ thereby constructed satisfies the constraints it is then accepted and we continue to the next iteration; otherwise we set
\begin{equation}
\label{nr correction}
\boldsymbol{\theta}^{(n,\,2)} :=\boldsymbol{\theta}^{(n,\,1)}+ \frac{\boldsymbol{\theta}^{(n,\,2)} -\boldsymbol{\theta}^{(n,\,1)} }{2} 
\end{equation}
iteratively until $\boldsymbol{\theta}^{(n,\,2)} $ satisfies the constraints. Notice that by repeatedly performing~(\ref{nr correction}), the value $\boldsymbol{\theta}^{(n,\,2)} $ hence constructed is closer and closer to $\boldsymbol{\theta}^{(n,\,1)} $; since this latter point satisfies the constraints, so will at some point $\boldsymbol{\theta}^{(n,\,2)} $. This manipulation that took us from $\boldsymbol{\theta}^{(n,\,1)}$ to $\boldsymbol{\theta}^{(n,\,2)} $ in a constraint compliant fashion has to be carried out at each iteration to go from $\boldsymbol{\theta}^{(n,\,k)}$ to $\boldsymbol{\theta}^{(n,\,k+1)} $.

\medskip

\noindent {\bf Trust-region iteration acceptance correction:} given an starting point $\boldsymbol{\theta}^0 $ we have given a prescription for the construction of a sequence $\{ \boldsymbol{\theta}^{(n)}\} _{n \in \mathbb{N}}$ that converges to the constrained minimizer of minus the quasi-loglikelihood $f(\boldsymbol{\theta}):=-{\rm log}L({\bf z}; \boldsymbol{\theta}) $. We now couple this optimization routine with a trust-region technique. The trust-region algorithm provides us with a systematic method to test the pertinence of an iteration before it is accepted and to adaptively modify the strength of the local penalization in order to speed up the convergence speed. In order to carefully explain our use of this procedure consider first the local model~(\ref{local optimization problem}) in which all the constants $L _1, \ldots, L _6 $ that manage the strength of the constraint penalizations are set to a common value $L$. At each iteration of~(\ref{local optimization problem bfgs}) compute the {\bf adequacy ratio} $\rho ^{(n)} $ defined as
\begin{equation}
\label{adequacy ratio}
\rho ^{(n)}:= \frac{f(\boldsymbol{\theta}^{(n)})-f(\boldsymbol{\theta}^{(n-1)})}{\hat{f} ^{(n)}(\boldsymbol{\theta}^{(n)})-\hat{f}^{(n)}(\boldsymbol{\theta}^{(n-1)})}
\end{equation}
which measures how close the descent in the target function in the present iteration is to the one exhibited by the local model $\hat{f}^{(n)} $. The values that can be obtained for $\rho^{(n)} $ are classified into three categories that determine different courses of action:
\begin{enumerate}
\item {\bf Too large step} $\rho ^{(n)}<0.01 $: there is too much dissimilarity between the local penalized model and the actual target function. In this situation, the iteration update is rejected by setting $\boldsymbol{\theta}^{(n)}=\boldsymbol{\theta}^{(n-1)}$ and the penalization is strengthened by doubling the constant: $L=2L  $
\item {\bf Good step} $0.01\leq\rho ^{(n)}\leq 0.9 $: the iteration update is accepted and the constant $L$ is left unchanged.
\item {\bf Too small step} $0.9\leq \rho ^{(n)}$: the iteration update is accepted but given the very good adequacy between the local penalized model and the target function we can afford loosening the penalization strength by setting $L= \frac{1}{2}L  $ as the constant that will be used in the next iteration.
\end{enumerate} 

\begin{remark}
\normalfont
Even though the definition of the adequacy ratio in~(\ref{adequacy ratio}) uses the full penalized local models $\hat{f}^{(n)} $, we have seen that in practice the linear approximation suffices to obtain good results. 
\end{remark}

\subsection{Preliminary estimation}
\label{Preliminary estimation}

As any optimization algorithm, the one that we just presented  requires a starting point $\boldsymbol{\theta}^{(0)} $. The choice of a good preliminary estimation of $\boldsymbol{\theta}^{(0)} $ is particularly relevant in our situation since the quasi-loglikelihood exhibits generically local extrema and hence initializing the optimization algorithm close enough to the solution may prove to be crucial in order to obtain the correct solution. 

Given a sample ${\bf z}=\{ {\bf z}_1, \ldots, {\bf z}_T \} $, a reasonable estimation for $\boldsymbol{\theta}^{(0)} $ can be obtained by using the following two steps scheme:

\medskip

\noindent {\bf 1. Find a preliminary estimation of the conditional covariance matrices sequence} $\{H _1, \ldots, H _T\} $ out of the sample ${\bf z} $. This can be achieved by using a variety of existing non-computationally intensive techniques. A non-exhaustive list is:
\begin{description}
\item [(i)] Orthogonal GARCH model (O-GARCH): introduced in~\cite{ding:thesis, alexander:chibumba, alexander:ogarch, alexander:covariance_matrices}; this technique is based on fitting one-dimensional GARCH models to the principal components obtained out of the sample marginal covariance matrix of ${\bf z}$.
\item [(ii)] Generalized orthogonal GARCH model (GO-GARCH)~\cite{gogarch}: similar to  O-GARCH, but in this case the one-dimensional modeling is carried out not for the principal components of ${\bf z} $ but for its image with respect to a transformation $V$ which is assumed is assumed to be just invertible (in the case of O-GARCH is also orthogonal) and it is estimated directly via a maximum likelihood procedure, together with the parameters of the one-dimensional GARCH models. GO-GARCH produces better empirical results than O-GARCH but it lacks the factoring estimation feature that O-GARCH has, making it more complicated for the modeling of large dimensional time series and conditional covariance matrices. 
\item [(iii)] Independent component analysis (ICA-GARCH):~\cite{wu:PCA, pena:ica} this model is based on a signal separation technique~\cite{comon:ica, fastica} that turns the time series into statistically independent components that are then treated separately using one dimensional GARCH models.
\item [(iv)] Dynamic conditional correlation model (DCC): introduced in~\cite{tse:dcc, engle:dcc}, this model proposes a dynamic behavior of the conditional correlation that depends on a small number of parameters and that nevertheless is still capable of capturing some of the features of more complicated multivariate models. Moreover, a version of this model~\cite{engle:sheppard:dcc} can be estimated consistently using a two-step approach that makes it suitable to handle large dimensional problems.
\end{description}
Another method that is widely used in the context of financial log-returns is the one advocated by Riskmetrics~\cite{riskmetrics} that proposes exponentially weighted moving average (EWMA) models for the time evolution of variances and covariances; this comes down to working with IGARCH type models with a coefficient that is not estimated but proposed by Riskmetrics and that is the same for all the factors.

\medskip

\noindent {\bf 2. Estimation of $\boldsymbol{\theta}^{(0)} $ out of ${\bf z} $ and $H=\{H _t\}_{t \in \{1, \ldots, T\}} $ using constrained ordinary least squares.} If we have the sample ${\bf z} $ and a preliminary estimation of the conditional covariances $\{H _t\}_{t \in \{1, \ldots, T\}} $, a good candidate for $\boldsymbol{\theta}^{(0)}=(A^{(0)},B^{(0)}, {\bf c}^{(0)}) $ is the value that minimizes the sum of the Euclidean norms $s _t:=\|{\bf h} _t-  \left( {\bf c}+A \boldsymbol{\eta}_{t-1}+B {\bf h}_{t-1}\right)\| ^2 $, that is,
\begin{equation*}
s(A,B, {\bf c}; {\bf z}, H)=\sum_{t=2}^T s_t(A,B, {\bf c}; {\bf z}, H)=\sum_{t=2}^T \|{\bf h} _t-  \left( {\bf c}+A \boldsymbol{\eta}_{t-1}+B {\bf h}_{t-1}\right)\| ^2,
\end{equation*}
subjected to the constraints {\bf (SC)}, {\bf (PC)}, {\bf (CC)}, and {\bf (KC)}. This minimizer can be efficiently found by using the Bregman divergences based method introduced in Sections~\ref{Constrained optimization via Bregman divergences} through~\ref{Performance improvement: BFGS and trust-region corrections} with the function $s(A,B, {\bf c}; {\bf z}, H) $ replacing minus the log-likelihood. However, we emphasize that unlike the situation in the log-likelihood problem, the choice of a starting point in the optimization of $s(A,B, {\bf c}; {\bf z}, H) $ is irrelevant given the convexity of his function.

As a consequence of these arguments, the preliminary estimation $\boldsymbol{\theta}^{(0)} $ is obtained by iterating~(\ref{local optimization problem bfgs}) where in the local model~(\ref{definition local model}) the map $f$ is replaced by $s$. This scheme is hence readily applicable once the gradient of $s$, provided by the following formulas, is available:
\begin{eqnarray*}
\nabla_A s&=&2\sum_{t=2}^T\left[A \boldsymbol{\eta}_{t-1}\boldsymbol{\eta}_{t-1}^T+ {\bf c} \boldsymbol{\eta}_{t-1}^T+B {\bf h}_{t-1}\boldsymbol{\eta}_{t-1} ^T- {\bf h}_t\boldsymbol{\eta}_{t-1} ^T\right],\\
\nabla_B s&=&2\sum_{t=2}^T\left[{\bf c}{\bf h}_{t-1}^T+A \boldsymbol{\eta}_{t-1} {\bf h}_{t-1} ^T+B {\bf h}_{t-1} {\bf h}_{t-1} ^T- {\bf h}_{t} {\bf h}_{t-1} ^T\right],\\
\nabla_{{\bf c}}s &= &2\sum_{t=2}^T \left[{\bf c}+A \boldsymbol{\eta}_{t-1}+B {\bf h}_{t-1}- {\bf h}_t\right].
\end{eqnarray*}

\section{Numerical experiments}
\label{numerical experiments}

In this section we illustrate the estimation method presented in Section~\ref{Calibration via Bregman matrix divergences} with various simulations that give an idea of the associated computational effort and of the pertinence of the VEC model in different dimensions.

\medskip

\begin{figure}[!htp]
\includegraphics[scale=.41]{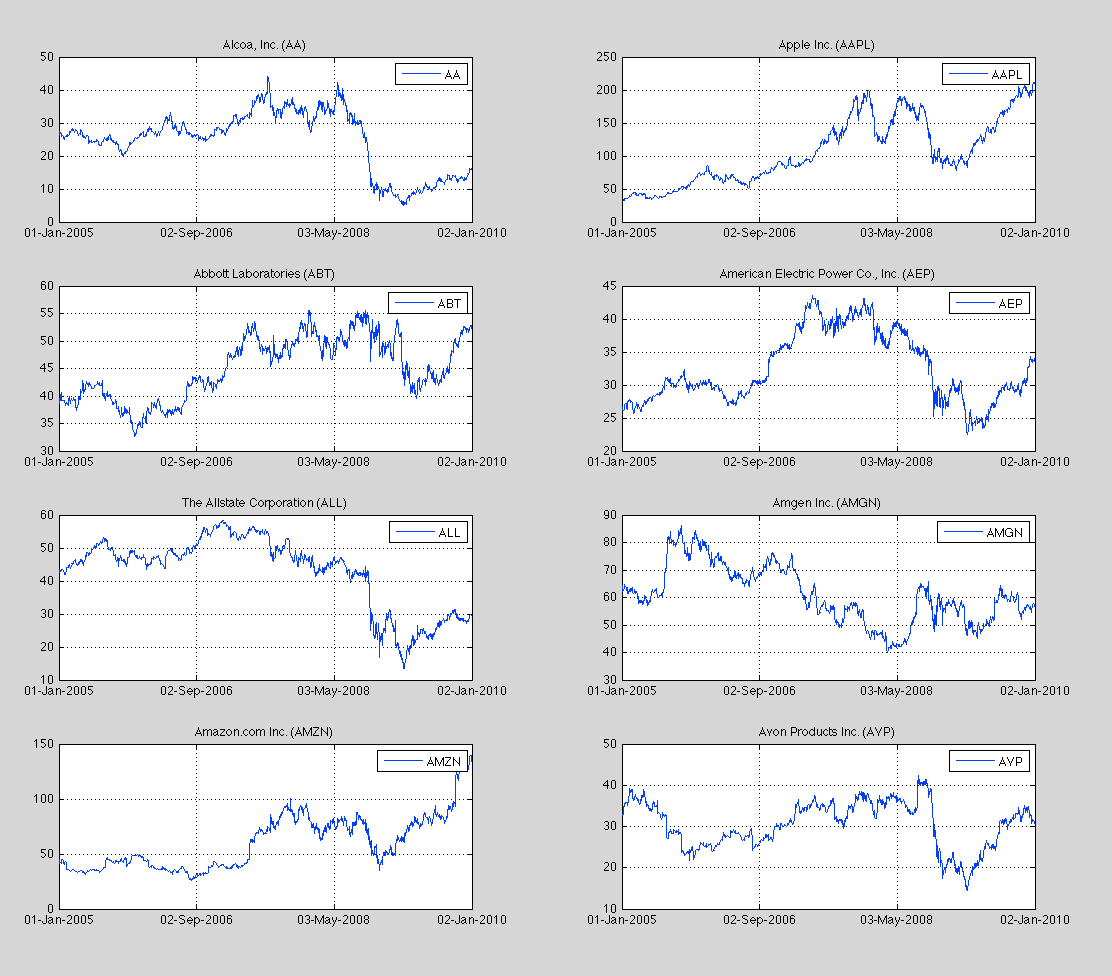}
\caption{Stock quotes used in the numerical experiments. The quotes represent closing prices adjusted with respect to dividend payments and stock splits. Source: Yahoo Finance.}
\label{fig:prices}
\end{figure}

\noindent {\bf The data set.} We have used in our experiments the daily closing prices between January 3, 2005 and December 31, 2009 (that is, 1258 date entries) of the stock associated to the companies Alcoa, Apple, Abbott Laboratories, American Electric, Allstate, Amgen, Amazon.com, and Avon. All these stocks are traded at the NYSE in US dollars and, in the last date of our sample, they were all constituents of the S\&P500 index. The quotes are adjusted with respect to dividend payments and stock splits. Figure~\ref{fig:prices} represents graphically the data set.

\medskip

\noindent {\bf Computational effort associated to the estimation method.} In table~\ref{fig:numerical performance} we have gathered the required computing time and the necessary gradient calls to fit VEC(1,1) models to the log-returns of our data set in different dimensions. In the $n=1 $ column we present the results associated to fitting a VEC model to the log-returns of the first element of the data set; the same in the $n=2$ column with respect to the log-returns of the first two elements of the data set, and so on. The stopping criterion for the algorithm is established by setting a termination tolerance on the function value equal to $10^{-5} $. The last row of the table shows how the algorithm becomes increasingly costlier with the dimensionality of the problem when the BFGS correction is dropped.The results of this experiment suggest that the trust-region correction speeds up the algorithm and the BFGS modification makes the convergence rate dimensionally independent.

\begin{table}[!htb]
\vspace{1.5ex}
\noindent\makebox[\textwidth]{%
\begin{tabularx}{1.25\textwidth}{X}
\scalebox{.70}{
\begin{tabular}{l*{2}{c}*{2}{c}*{2}{c}*{2}{c}*{2}{c}*{2}{c}}
\toprule
\multicolumn{13}{c}{{\bf Computation time and gradient calls}} \\
\cmidrule(r){2-13}
	&\multicolumn{2}{c}{$n=1$ (3 parameters)}&\multicolumn{2}{c}{$n=2$ (21 parameters)}&\multicolumn{2}{c}{$n=3$ (78 parameters)}&\multicolumn{2}{c}{$n=4$ (210 parameters)}&\multicolumn{2}{c}{$n=5$ (465 parameters)}&\multicolumn{2}{c}{$n=6$ (903 parameters)}\\
\cmidrule(r){2-3}\cmidrule(r){4-5}\cmidrule(r){6-7}\cmidrule(r){8-9}\cmidrule(r){10-11}\cmidrule(r){12-13}
	& Grad. calls &Time& Grad. calls &Time& Grad. calls &Time& Grad. calls &Time& Grad. calls &Time& Grad. calls &Time\\
\midrule
{\bf Full method}& 50 &1.45 sec &97 &58 sec &99 &3 min 9 sec &94 &18 min &85 &73 min &105 & 5 hrs 36 min\\
\midrule
{\bf No BFGS}& 106 &2.30 sec &281 &159 sec &378  &8 min 57 sec&404 &47 min &534 &298 min &591 &22 hr\\
\bottomrule
\end{tabular}}
\end{tabularx}}
\caption{Computation time and gradient calls required when running the estimation method presented in Section~\ref{Calibration via Bregman matrix divergences}, with and without the BFGS correction. The simulations were carried out using a nonparallelized Matlab script on an Apple computer endowed with two double core 3 GHz processors, 64 bits.}
\label{fig:numerical performance}
\end{table}

\medskip

\noindent {\bf Variance minimizing portfolios, proxy replication, and spectral sparsity.} As we have already pointed out several times, the main concern when using VEC models lays in the overabundance of parameters, whose number may easily be bigger than the sample size in standard applications, even when dealing with low dimensional problems. This lack of parsimony already appears when dealing with our data set for it contains 1257 historical log-returns, while the VEC(1,1) model requires 1596 parameters in dimension seven and 2628 in dimension 8.

The goal of the following experiment consists of assessing how serious this problem is. More explicitly, we will study how the pertinence of VEC as a modeling tool evolves with the increase in dimensionality, when compared with other more parsimonious and widely used alternatives, namely:
\begin{itemize}
\item Exponentially Weighted Moving Average (EWMA) model for the conditional covariance matrices with the autroregressive coefficient $ \lambda=0.94 $ proposed by Riskmetrics~\cite{riskmetrics} for daily data.
\item Orthogonal GARCH model (OGARCH), as in~\cite{ding:thesis, alexander:chibumba, alexander:ogarch, alexander:covariance_matrices}.
\item Dynamic conditional correlation model (DCC) of~\cite{tse:dcc, engle:dcc}.
\end{itemize}
These modeling approaches will be tested by evaluating:
\begin{itemize}
\item {\bf Comparative performance in the construction of dynamic variance minimizing portfolios:} all the models that we just enumerated and that take part in our comparison share the form
\begin{equation}
\label{model for logreturns}
\mathbf{z}_t = H_t^{1/2} \boldsymbol{\epsilon _t} \quad\quad\text{with}\quad\quad
\{\boldsymbol{\epsilon _t}\}\sim {\rm IIDN}({\boldsymbol 0}, {\boldsymbol I}_n),
\end{equation}
where $\{H_t\} $ is a predictable matrix process. What changes from model to model is the specification that determines the dynamical behavior of $\{H_t\} $; in the particular case of {\rm VEC}(1,1), that specification is spelled out in~(\ref{vec11 model}). When~(\ref{model for logreturns}) is fitted to the log-returns associated to our data set,  the matrices $\{H_t\} $ provide an (model dependent) estimate of the conditional covariance of the log-returns process. Moreover, it is not difficult to show that if  $\mathbf{w}=(w _1, \ldots , w _n)' $ is a weights vector such that  $\sum_{i=1}^n w _i=1 $,  then the conditional variance of the net returns process of the associated portfolio is given by $\{\mathbf{w}^TA _t \mathbf{w}\} $, where $A _t $ is the matrix whose  $(i,j)$ entry $A _{ij}^t $ is given by
\begin{equation*}
A _{ij}^t=\exp \left(\sum _{k=1}^n h_{ik}h_{jk}\right)-1,
\end{equation*}
with $h ^t_{ij} $ the $(i,j)$ entry of the matrix $H _t $. A dynamic variance minimizing portfolio is a weights vector $\mathbf{w}_t $ defined as the solution at each time step of the optimization problem
\begin{equation}
\label{dynamic variance minimizing portfolio}
\mathop{\rm arg\, min}_{\mathbf{w}\in \mathbb{R}^n, \, \sum_{i=1}^n w _i=1}\, \mathbf{w}^TA _t \mathbf{w}. 
\end{equation}
A straightforward application of Lagrange duality shows that the solutions $\mathbf{w}_t $ of~(\ref{dynamic variance minimizing portfolio}) are given by either the zero eigenvectors of $A _t $, or by
\begin{equation}
\label{solution variance minimizing}
\mathbf{w} _t= \frac{1}{\mathbf{i}^T A_t ^{-1} \mathbf{i}} A_t ^{-1} \mathbf{i},
\end{equation}
when $A _t $ is invertible, where $\mathbf{i} $ is an $n $-dimensional vector made exclusively out of ones. In our numerical experiment we will always fall in the situation contemplated in~(\ref{solution variance minimizing}) and it is this expression that we will use to construct the dynamic variance minimizing portfolios associated to each of the different models that we are testing. Figure~\ref{fig:variance_minimizing} shows the conditional variance of the net returns process associated to the variance minimizing portfolios corresponding to the different models under consideration. It is tempting to say that the most performing model is the one for which the conditional variance is consistently smaller; however, given that the conditional variance is model dependent, {\it these quantities are not directly comparable} and it is only the marginal variances of the optimal portfolios that can be put side to side. This comparison is carried out in table~\ref{fig:Variance of optimal portfolios table} in which we see that VEC allows the construction of portfolios with smaller variances than those corresponding to the other models in all the dimensions considered.

\medskip

\begin{figure}[htp]
\includegraphics[scale=.41]{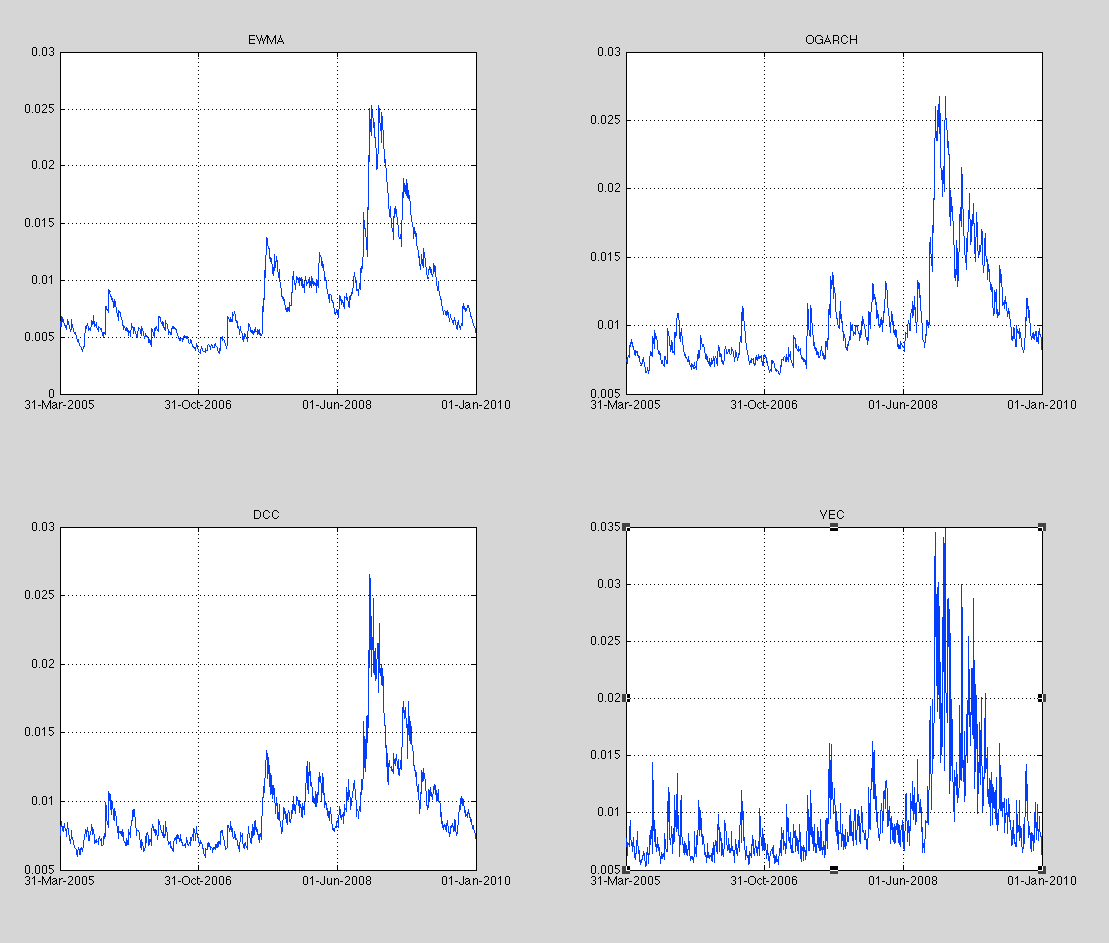}
\caption{Conditional volatility of the net returns associated to the dynamic variance minimizing portfolios constructed by fitting different models to the log-returns of our eight dimensional data set. The VEC estimation was carried out setting a termination tolerance on the function value equal to $10^{-5} $ and using an OGARCH based OLS preliminary estimation, as explained in Section~\ref{Preliminary estimation}.}
\label{fig:variance_minimizing}
\end{figure}

\medskip 

\begin{table}
\noindent\makebox[\textwidth]{%
\begin{tabularx}{.85\textwidth}{X}
\scalebox{.85}{
\begin{tabular}{l*{7}{c}}
\toprule
	&\multicolumn{7}{c}{{\bf Variance of optimal portfolios ($\times 10^{-4}$)}} \\
\cmidrule(r){2-8}
	&$n=2 $&$n=3 $&$n=4 $&$n=5 $&$n=6 $&$n=7 $&$n=8$\\
\midrule
{\bf EWMA} &5.03 &1.72 &1.50 &1.44 &1.49 & 1.59 & 1.62\\
\midrule
{\bf OGARCH} &5.29 &1.75 & 1.50 & 1.42& 1.42 &1.45 &1.50\\
\midrule
{\bf DCC} &4.96 & 1.74 &1.47 &1.37 &1.38 &1.40 &1.43\\
\midrule
{\bf VEC} &\bf 4.91 &\bf 1.70 &\bf 1.39 &\bf 1.22 &\bf 1.15 &\bf 1.15 &\bf 1.12\\
\bottomrule
\end{tabular}}
\end{tabularx}}
\caption{Marginal variance of the net returns associated to the variance minimizing portfolios corresponding to the different models.}
\label{fig:Variance of optimal portfolios table}
\end{table}

\item {\bf Goodness of fit between the associated conditional volatilities and the absolute values of the log-returns used as a proxy for conditional volatility:} following~\cite{mincer:zarnowitz, granger:newbold, andersen:bollerslev, manganelli:sav:ecb}, we evaluate the performance of the different models by considering the absolute values of portfolio returns as a proxy for conditional volatility and by checking how the different proposals coming from the models under scrutiny fit this proxy. Even though it is well known~\cite{andersen:bollerslev} that this is a very noisy proxy for volatility, this approach provides us with quick and simple to implement ways to compare different modeling approaches. The first one consists of fitting each of the models to the first $i$ assets with $i\in \left\{1, 2, \ldots ,8\right\}$ and computing the mean Euclidean distance (MSE) between the model associated conditional volatility and the proxy values; the results of this experiment are presented in table~\ref{fig:proxy errors table} where we see that VEC produces a smaller MSE in all the dimensions considered, even surprisingly at dimensions $7$ and $8$ where the number of parameters to be estimated is bigger than the sample size. The plausibility of this result is visually emphasized in figure~\ref{fig:volatilities} where we have depicted the conditional volatilities of one of the assets in our data set (AA) obtained out of the models under consideration in dimension $8$, as well as of a one-dimensional GARCH model; these volatilities are graphically compared with the proxy.

\medskip

\begin{table}
\noindent\makebox[\textwidth]{%
\begin{tabularx}{.85\textwidth}{X}
\scalebox{.85}{
\begin{tabular}{l*{8}{c}}
\toprule
	&\multicolumn{8}{c}{{\bf Mean square error with respect to proxy ($\times 10^{-4}$)}} \\
\cmidrule(r){2-9}
	&$n=1 $&$n=2 $&$n=3 $&$n=4 $&$n=5 $&$n=6 $&$n=7 $&$n=8 $\\
\midrule
{\bf EWMA} &5.19 &4.31 &3.23 &2.71 &3.03 &2.89 &3.42 &3.39\\
\midrule
{\bf OGARCH}  &\bf 5.08 &4.33 &3.24 & 2.71 &3.02 &2.88 & 3.44 &3.38\\
\midrule
{\bf DCC} &\bf 5.08 &4.23 &3.17 &2.66 &2.98 &2.85 &3.39 &3.43\\
\midrule
{\bf VEC} &\bf 5.08 &\bf 4.17 &\bf 3.12 &\bf \bf 2.59 &\bf 2.91 &\bf 2.68 &\bf 3.17 &\bf 3.16\\
\bottomrule
\end{tabular}}
\end{tabularx}}
\caption{Mean square errors committed when modeling the absolute values of the returns with the conditional variance associated to the different models.}
\label{fig:proxy errors table}
\end{table}

\begin{figure}[htp]
\includegraphics[scale=.41]{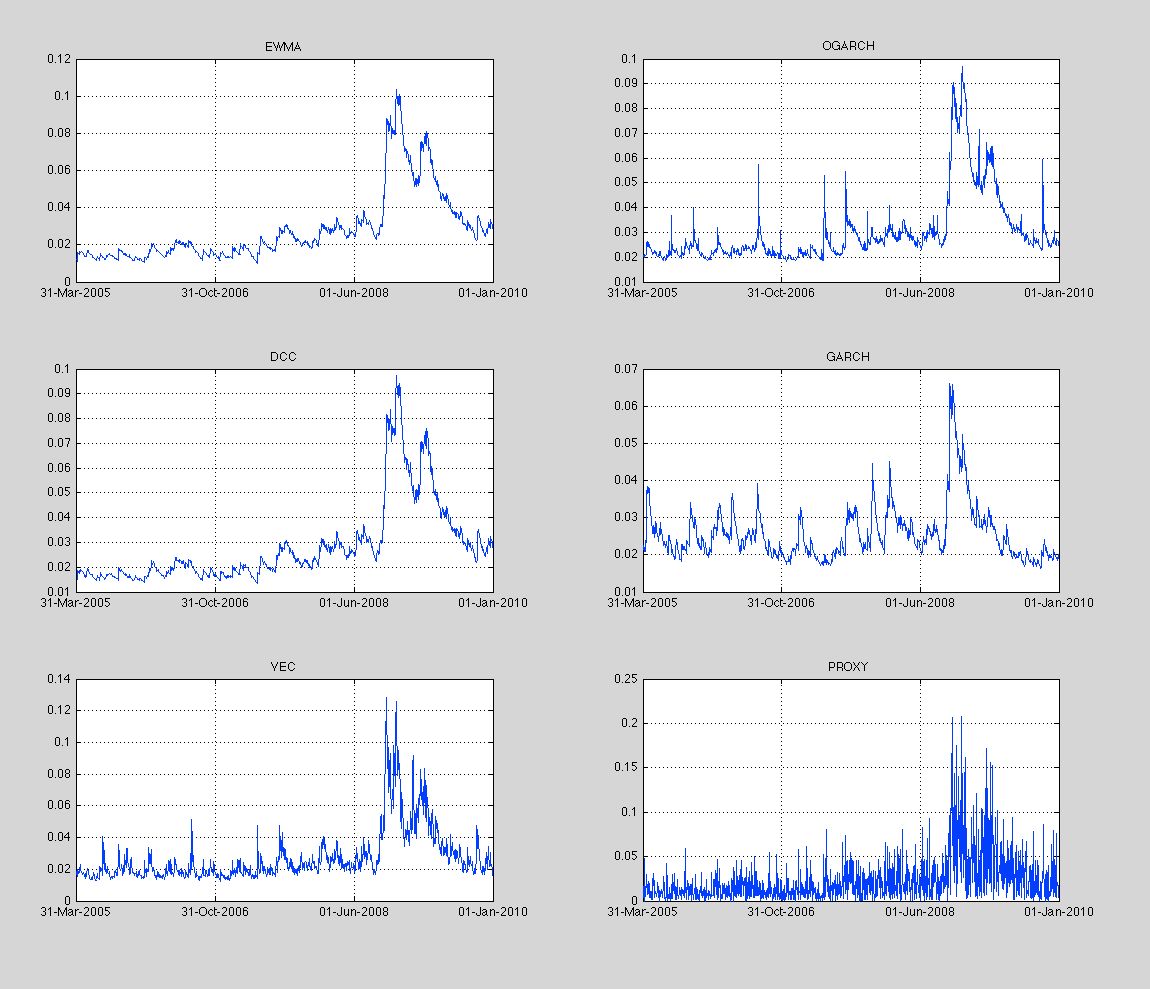}
\caption{Conditional volatility of the asset Alcoa (AA) obtained out of eight dimensional modelings. The graphics EWMA, OGARCH, DCC, and VEC represent the volatility obtained as the square root of the  $(1,1)$ components of the $8 \times 8 $ conditional covariance matrices associated to those models. GARCH represents the volatility associated to a one-dimensional GARCH modeling of the log-returns of AA, and PROXY shows the absolute value of the AA log-returns.}
\label{fig:volatilities}
\end{figure}

Finally, we have ranked the different models by studying their efficiency in modeling the volatility of constant random portfolios; more especifically, at each dimension $i$, $i\in \left\{1, 2, \ldots ,8\right\}$, we randomly choose $i$ weights using standard normally distributed variables and we appropriately normalize them so that their sum equals to one. We then use the conditional covariance matrices provided by each of the models under consideration to compute the (model based) conditional volatility of the portfolio. We then regress the proxy for the portfolio volatility, namely the absolute value of the portfolio returns, on the various portfolio volatilities provided by the different models and, using the suggestion in~\cite{manganelli:sav:ecb} we declare  as the best model the one that produces the highest coefficient of determination $R^2$. As the chosen proxy is know to be very noisy~\cite{andersen:bollerslev} the obtained $R ^2$ coefficients are rather small (typically between $0.2$ and $0.3 $). Using this criterion, we randomly generated 5,000 portfolios at each dimension, and we recorded the percentage rate of relative success of each model with respect to the others. The results of the experiment are presented in table~\ref{fig:Success rate in modeling} and show the superiority of VEC in all the dimensions considered. 

\medskip

\begin{table}[!htb]
\vspace{1.5ex}
\noindent\makebox[\textwidth]{%
\begin{tabularx}{0.85\textwidth}{X}
\scalebox{.85}{
\begin{tabular}{l*{7}{c}}
\toprule
	&\multicolumn{7}{c}{\bf Success rate in modeling random portfolios (\%)} \\
\cmidrule(r){2-8}
{\bf Number of assets}&n=8 &n=7 &n=6 &n=5 &n=4 &n=3 &n=2\\
\cmidrule(r){2-8}
{\bf Number of VEC parameters}&2628 &1596 &903 & 465 &210 &78 &21 \\
\midrule
{\bf DCC}&25.70 &32.20 &32.00 & 32.27 &17.70 &24.10 & 24.50\\
\midrule
{\bf EWMA}&1.27 &0.9 &0 &0 & 0 &0.3 & 0\\
\midrule
{\bf OGARCH}&28.18 &19.40 &8.30 &7.87 &3.70 &21.80 &22.80\\
\midrule
{\bf VEC} & {\bf 44.85} & {\bf 47.50} & {\bf 59.70} & {\bf 59.83} & {\bf 78.60} & {\bf 53.80} & {\bf 52.70}\\
\bottomrule
\end{tabular}}
\end{tabularx}}
\caption{Percentage rate of relative success of each model with respect to the others in modeling the volatility of random portfolios. For each dimension $n$, we randomly generated 5,000 portfolios and we considered as the best model the one that produced the highest coefficient of determination $R^2$ when regressing the absolute values of the corresponding returns on the conditional covariances associated to the each model.  VEC consitently presents the highest success rate regardless of the dimension.}
\label{fig:Success rate in modeling}
\end{table}

\item {\bf Spectral sparsity and high dimensional estimation:} a major surprise revealed by these numerical experiments is that the estimated models provide good empirical  performance despite the highly unfavorable ratio between the sample size and the number of parameters to be estimated. In order to investigate the reasons for such a counterintuitive but pleasing phenomenon, we plotted the eigenvalues of $\Sigma(A)$ and $\Sigma(B)$ for $n$ between 4 to 8. These plots, displayed in 
Figure \ref{fig:eigenvalues_sigma}, show that the estimators $\Sigma(\hat{A})$ and $\Sigma(\hat{B})$ of  $\Sigma(A)$ and $\Sigma(B)$ are spectrally very sparse, that is, have a very low rank. Thus, the solutions of the estimation problem are exactly the same as the ones we would have obtained under additional a priori rank constraints, a setting that would have  implicitly reduced the dimension of the parameter space by a large factor. This suggests that in our particular empirical situation, the number of parameters that  are actually independent is much smaller than the number of entries in the coefficient matrices  $A$, $B$ and  ${\bf c}$, which makes possible the  use of small relative sample sizes in the VEC context. The most obvious explanation for this phenomenon stems from the well-known fact that the conditional covariance matrices $H_t$ corresponding to stock market returns present spectral accumulation (in small dimensional settings) or sparsity (in large dimensions). This  seems to make the positive semi-definiteness constraints on $\Sigma(A)$ and $\Sigma(B)$ highly active, which enforces a large number of eigenvalues to be equal to zero. 

Notice further that the proportion of nonzero eigenvalues of
$\Sigma(\hat{A})$ and $\Sigma(\hat{B})$ decreases very slowly as a
function of the parameter space dimension and shows no particular
abrupt transition when the dimension/sample size ratio becomes large.
This phenomenon suggests that the constrained maximum likelihood
approach is very stable for this hard estimation problem. On the other
hand, pushing theses ideas further along the lines of recent works in
sparse estimation and matrix completion problems
\cite{CandesRecht:FondCompMath09, CandesTao:IEEEIT10}, one might
expect that explicitly enforcing the spectral sparsity of the
estimators might improve their performance  for
dimensions much larger than the ones explored in the present work.
A rigorous treatment of these observations is needed and
will be the subject of further research in a forthcoming paper.

\begin{figure}[htp]
\includegraphics[width=6.5in,height=7in]{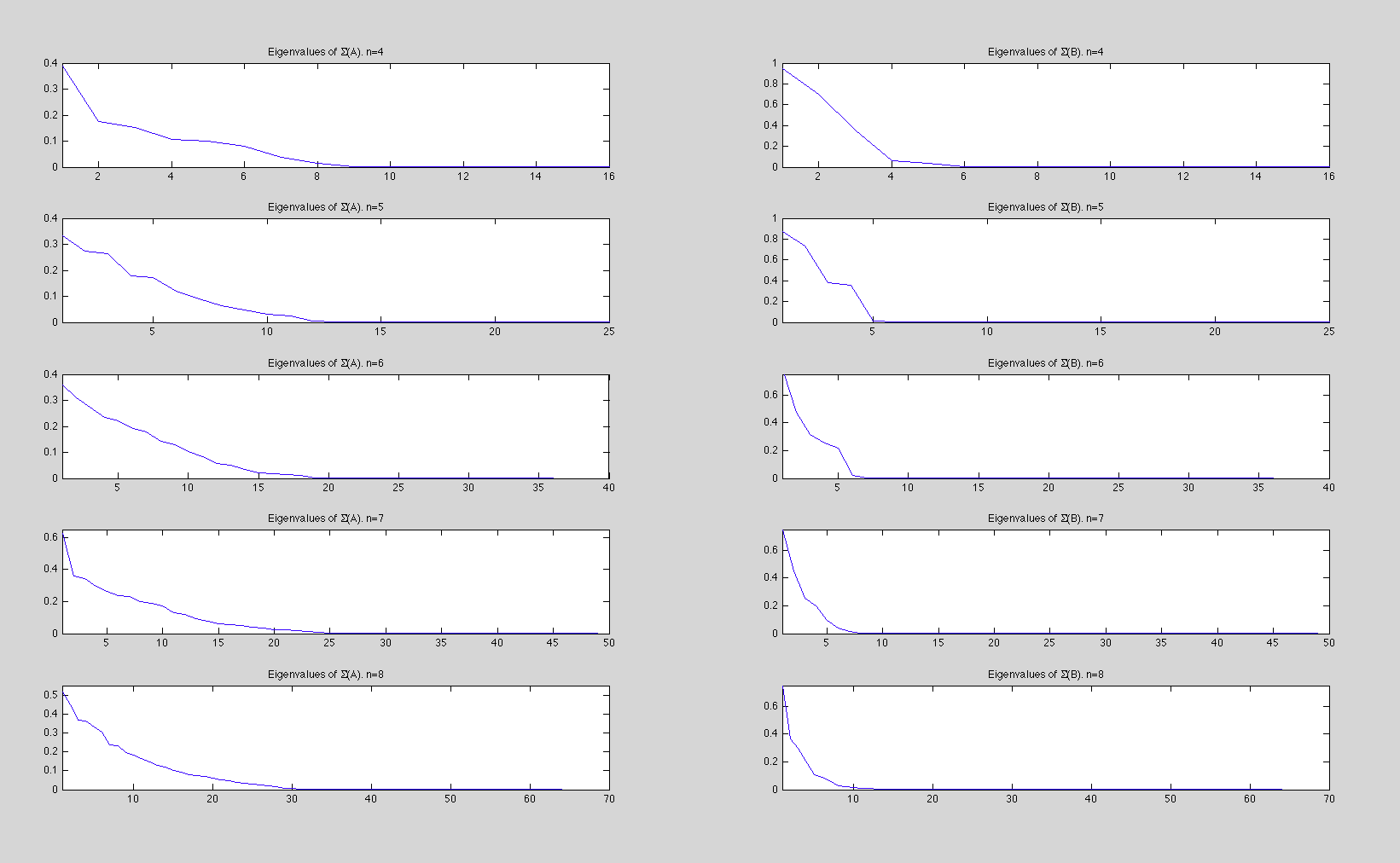}
\caption{Eigenvalues of $\Sigma(A)$ and $\Sigma(B)$ for $n$ between 4 to 8. The spectral sparsity evidenced in these plots suggests nonlinear constraints in the parameter space which explain the good empirical  performance of the models despite the highly unfavorable ratio between the sample size and the number of parameters to be estimated.}
\label{fig:eigenvalues_sigma}
\end{figure}
\end{itemize}

\section{Conclusions}

In this paper we provided an adequate explicit formulation of the  estimation problem for VEC models and developed a Bregman-proximal trust-region method to solve it. This combination of techniques provides a robust optimization method that can be surely adapted with good results to more parsimonious multivariate volatility models.

We carried out numerical experiments based on stock market returns that show the applicability of the proposed estimation method in specific practical situations. Additionally, our numerical experiments reveal how the empirically well documented spectral accumulation in the covariance structure of stock quotes implies, in the context of VEC modeling, implicit nonlinear constraints in the parameter space that make this parametric family competitive even in the presence of a highly unfavorable ratio between the sample size and the number of parameters to be estimated. The comparison has been carried out with respect to other standard and more parsimonious multivariate conditionally heteroscedastic families, namely, EWMA, DCC, and OGARCH. An in-depth study of this phenomenon will be the subject of a forthcoming publication.

\section{Appendix}
\subsection{Proof of Proposition~\ref{vec identities}}
We start with the proof of {\bf (i)} by using the following chain of equalities in which we use the symmetric character of both $A$ and ${\rm math}(m)$:
\begin{eqnarray*}
\langle A+ {\rm diag}(A), {\rm math}(m) \rangle &=&  \operatorname{trace}(A\,  {\rm math}(m))+ \operatorname{trace}({\rm diag}(A){\rm math}(m))\\
	&=&\sum_{i,j=1}^n A_{ij}{\rm math}(m)_{ji}+A_{ij}\delta _{ij}{\rm math}(m)_{ji}\\
	&=&\sum_{i<j} A_{ij}{\rm math}(m)_{ij}+\sum_{i>j} A_{ij}{\rm math}(m)_{ij}+2 \sum_{i=j=1}^n A_{ij}{\rm math}(m)_{ij}\\
	&= &2 \sum_{i\geq j} A_{ij}{\rm math}(m)_{ij}=2 \sum_{i\geq j} A_{ij}m_{\sigma(i,j)}=2 \sum_{q=1}^N A _{\sigma^{-1}(q)} m _q\\
	&= &2\langle {\rm vech}(A),m\rangle, 
\end{eqnarray*}
as required. In order to prove {\bf (ii)}, note that the identity that we just showed ensures that
\begin{equation}
\label{intermediate vecid}
\langle A, {\rm math}(m)\rangle=2 \langle {\rm vech}(A),m\rangle- \langle{\rm diag}(A), {\rm math}(m)\rangle. 
\end{equation}
At the same time
\begin{eqnarray*}
\langle{\rm diag}(A), {\rm math}(m)\rangle&= &\operatorname{trace}({\rm diag}(A){\rm math}(m))=\sum_{i=1}^n A_{ii} {\rm math}(m)_{ii}=\sum_{i=1}^n A_{ii} m_{\sigma(i,i)}\\
	&=&\sum_{i\geq j}{\rm diag}(A)_{ij} m_{\sigma(i,j)}=\sum_{q=1}^N{\rm diag}(A)_{\sigma^{-1}(q)} m_{q}\\
	&=&\sum_{q=1}^N{\rm vech}({\rm diag}(A))_{q} m_{q}= \langle{\rm vech}({\rm diag}(A)),m\rangle,
\end{eqnarray*}
which substituted in the right hand side of~(\ref{intermediate vecid}) proves the required identity. Finally, expression~(\ref{expression maths}) follows directly from {\bf (ii)} and as to~(\ref{expression vechs}) we observe that
\begin{eqnarray*}
\frac{1}{2}\langle A+ {\rm diag}(A), {\rm math}(m) \rangle &=& \frac{1}{2}\operatorname{trace}((A+ {\rm diag}(A)){\rm math}(m))\\
&=& \frac{1}{2}\operatorname{trace}(A {\rm math}(m))+ \frac{1}{2}\operatorname{trace}({\rm diag}(A) {\rm math}(m))\\
	&= &\frac{1}{2}\left(\operatorname{trace}(A {\rm math}(m))+\operatorname{trace}(A{\rm diag}( {\rm math}(m)))\right)\\
	&=& \frac{1}{2} \langle A, {\rm math}(m)+ {\rm diag}({\rm math}(m))\rangle,
\end{eqnarray*} 
which proves~(\ref{expression vechs}). Regarding the operator norms we will just prove~(\ref{operator norm vech}) and~(\ref{operator norm math}) as the rest can be easily obtained out of these two combined with the expressions~(\ref{expression maths}) and~(\ref{expression vechs}). We start by noticing that for any nonzero $A=(a_{ij}) \in \Bbb S_n $:
\begin{equation*}
\frac{\|{\rm vech}(A)\| ^2}{\|A\| ^2}= \frac{\sum_{i>j=1}^n a_{ij}^2+\sum_{i=1}^n a_{ii}^2}{2\sum_{i>j=1}^n a_{ij}^2+\sum_{i=1}^n a_{ii}^2}=1-\frac{\sum_{i>j=1}^n a_{ij}^2}{2\sum_{i>j=1}^n a_{ij}^2+\sum_{i=1}^n a_{ii}^2}.
\end{equation*}
Since the last summand in the previous expression is always positive we have that
\begin{equation*}
\| {\rm vech}\|_{op}=\sup_{A \in \Bbb S _n, A\neq 0}\frac{\|{\rm vech}(A)\| }{\|A\| }=1,
\end{equation*}
the supremum being attained by any diagonal matrix ($\sum_{i>j=1}^n a_{ij}^2=0 $ in that case). Consider now $v= {\rm vech}(A)$. Then:
\begin{equation}
\label{from a to v}
\frac{\| {\rm math}(v)\| ^2}{\|v\| ^2}= \frac{\|A\| ^2}{\|{\rm vech}(A)\| ^2}= \frac{2\sum_{i>j=1}^n a_{ij}^2+\sum_{i=1}^n a_{ii}^2}{\sum_{i>j=1}^n a_{ij}^2+\sum_{i=1}^n a_{ii}^2}=1+\frac{\sum_{i>j=1}^n a_{ij}^2}{\sum_{i>j=1}^n a_{ij}^2+\sum_{i=1}^n a_{ii}^2}.
\end{equation}
When we let $A \in \Bbb S_n $ vary in the previous expression, we obtain a supremum by considering matrices with zeros in the diagonal ($\sum_{i=1}^n a_{ii}^2=0 $) and by choosing $\sum_{i>j=1}^n a_{ij}^2 \rightarrow \infty $, in which case $ \frac{\|A\| ^2}{\|{\rm vech}(A)\| ^2} \rightarrow 2 $.
Finally, as the map ${\rm vech}: \Bbb S _n\rightarrow \mathbb{R}^N$ is an isomorphism,~(\ref{from a to v}) implies that 
\begin{equation*}
\| {\rm math}\|_{op}=\sup_{v \in \mathbb{R}^N, v\neq 0} \frac{\| {\rm math}(v)\| }{\|v\| }= \sup_{A \in \Bbb S _n, A\neq 0} \frac{\|A\| }{\|{\rm vech}(A)\| }=\sqrt{2}.\quad  \blacksquare
\end{equation*}

\subsection{Proof of Proposition~\ref{property of sigmaa statement}}

We just need to verify that~(\ref{sigma definition}) satisfies~(\ref{property of sigmaa}). Let $k, l \in \{1, \ldots,n\} $ be such that $k\geq l $. Then,
\begin{eqnarray*}
(A \, {\rm vech}(H))_{\sigma(k,l)}&=& \sum_{i\geq j} A_{\sigma(k,l), \sigma(i,j)}H_{ij}= \sum_{i\geq j} A_{\sigma(k,l), \sigma(i,j)}\frac{H_{ij}+H_{ji}}{2}\\
	&=& \frac{1}{2}\sum_{i\geq j} A_{\sigma(k,l), \sigma(i,j)}H_{ij}+\frac{1}{2}\sum_{i\geq j} A_{\sigma(k,l), \sigma(i,j)}H_{ji}\\
	&=& \frac{1}{2}\sum_{i> j} A_{\sigma(k,l), \sigma(i,j)}H_{ij}+\sum_{i= j} A_{\sigma(k,l), \sigma(i,j)}H_{ij}+\frac{1}{2}\sum_{i< j} A_{\sigma(k,l), \sigma(j,i)}H_{ij}\\
	&=&\sum_{i> j}(\Sigma (A)_{kl})_{ij} H_{ij}+\sum_{i= j} (\Sigma (A)_{kl})_{ij} H_{ij}+\sum_{i<j}(\Sigma (A)_{kl})_{ij} H_{ij}=\operatorname{trace}(\Sigma (A)_{kl}H),
\end{eqnarray*} 
as required. \quad $\blacksquare$

\subsection{Proof of Proposition~\ref{sigma dual}}

We start with the following Lemma:

\begin{lemma}
\label{orthogonal projection n symmetric}
Let $A \in \mathbb{M}_{n ^2}$. The orthogonal projections $\mathbb{P}_{n ^2} (A) \in \Bbb S_{n ^2}$ and $\mathbb{P}_{n ^2}^n (A) \in \Bbb S_{n ^2}^n$ of $A$ onto the spaces of symmetric and $n$-symmetric matrices with respect to the Frobenius inner product~(\ref{frobenius inner product}) are given by:
\begin{eqnarray}
\mathbb{P}_{n ^2} (A) &= & \frac{1}{2}(A+A ^T)\label{symmetric 1}\\
(\mathbb{P}_{n ^2}^n (A))_{kl}&= & \frac{1}{4}(A_{kl}+A_{kl}^T+A_{lk}+A_{lk}^T),\label{symmetric 2}
\end{eqnarray}
for any block $(\mathbb{P}_{n ^2}^n (A))_{kl} $ of $\mathbb{P}_{n ^2}^n (A) $, $k,l \in \{1, \ldots, n\} $. 
\end{lemma}

\noindent\textbf{Proof.\ \ } In order to prove~(\ref{symmetric 1}) it suffices to check that $ \langle A-\mathbb{P}_{n ^2} (A), B\rangle=0 $ for any $B \in \Bbb S_{n ^2}$. Indeed,
\begin{equation*}
\langle A-\mathbb{P}_{n ^2} (A), B\rangle= \operatorname{trace}(AB)- \frac{1}{2}\operatorname{trace}(AB) - \frac{1}{2}  \operatorname{trace}(A^TB)=0.
\end{equation*} 
The result follows from the uniqueness of the orthogonal projection. Regarding  ~(\ref{symmetric 2}) we check that $ \langle A-\mathbb{P}_{n ^2}^n (A), B\rangle=0 $, for any $B \in \Bbb S_{n ^2}^n$. Given that for any $k,l \in \{1, \ldots, n\} $ the block $(AB)_{kl} $ is given by $(AB)_{kl}=\sum_{r=1}^n A_{kr}B_{rl}$ we have
\begin{eqnarray*}
\langle A-\mathbb{P}_{n ^2} ^n(A), B\rangle &= &\operatorname{trace}(AB)- \operatorname{trace}(\mathbb{P}_{n ^2} ^n(A)B)=\sum_{i=1}^n \operatorname{trace}(AB)_{ii}-\operatorname{trace}(\mathbb{P}_{n ^2} ^n(A)B)_{ii}\\
 	&= &\sum_{i,j=1}^n \operatorname{trace}(A_{ij}B_{ji})- \operatorname{trace}((\mathbb{P}_{n ^2} ^n(A))_{ij}B_{ji})=\sum_{i,j=1}^n \operatorname{trace}(A_{ij}B_{ji})\\
	& -&\sum_{i,j=1}^n \left[ \frac{1}{4}\operatorname{trace}(A_{ij}B_{ji})+ \frac{1}{4}\operatorname{trace}(A_{ij}^TB_{ji})+ \frac{1}{4}\operatorname{trace}(A_{ji}B_{ji})+\frac{1}{4}\operatorname{trace}(A_{ji}^TB_{ji})\right]=0,
\end{eqnarray*}
where we used that, due to the $n$-symmetricity of $B$ $\operatorname{trace}(A_{ij}^TB_{ji})=\operatorname{trace}(B_{ji}^TA_{ij})=\operatorname{trace}(A_{ij}B_{ji}) $ and
\begin{equation*}
\sum_{i,j=1}^n\operatorname{trace}(A_{ji}B_{ji})=\operatorname{trace}(A_{ji}B_{ij})=\operatorname{trace}(A_{ij}B_{ij}).
\end{equation*}
Analogously $\sum_{i,j=1}^n\operatorname{trace}(A_{ji}^TB_{ji})=\operatorname{trace}(A_{ij}B_{ij})$. \quad $\blacksquare$ 

\medskip

Now, in order to prove Proposition~\ref{sigma dual}, consider $A\in \mathbb{M}_N$  and $\mathcal{B} \in \mathbb{M}_{n ^2}$. Since the image of the map $\Sigma $ lies in $\Bbb S_{n ^2}^2 $ we have that $\langle\mathcal{B}-\mathbb{P}_{n ^2}^n(\mathcal{B}), \Sigma(A)\rangle=0 $ and hence
\begin{equation*}
\langle\Sigma^\ast (\mathcal{B}), A \rangle =  \langle \mathcal{B}, \Sigma(A)\rangle=\langle \mathbb{P}_{n ^2}^n(\mathcal{B})+\mathcal{B}-\mathbb{P}_{n ^2}^n(\mathcal{B}), \Sigma(A)\rangle=\langle \mathbb{P}_{n ^2}^n(\mathcal{B}), \Sigma(A)\rangle=\langle \Sigma^\ast (\mathbb{P}_{n ^2}^n(\mathcal{B})), A\rangle.
\end{equation*}
This identity allows us to restrict the proof of~(\ref{sigma dual}) to the $n$-symmetric elements $\mathcal{B}\in \Bbb S_{n^2} ^n  $. Hence let $\mathcal{B}\in \Bbb S_{n^2} ^n  $ and let $\widetilde{\sigma} $ be the extension of the map $\sigma$ defined in~(\ref{extension of sigma}). Then,
\begin{eqnarray*}
\langle\Sigma(A), \mathcal{B}\rangle &=&\sum_{k,l=1}^n \langle\Sigma(A)_{kl}, \mathcal{B} _{kl}\rangle=\sum_{k,l=1}^n \operatorname{trace}(\Sigma(A)_{kl}\mathcal{B} _{kl}^T)=\sum_{k,l,i,j=1}^n (\Sigma(A)_{kl})_{ij}(\mathcal{B} _{kl})_{ij}\\
	&= &\sum_{k,l,i,j=1}^n \frac{1}{2}\left[A_{\widetilde{\sigma}(k,l), \widetilde{\sigma}(i,j)} +A_{\widetilde{\sigma}(k,l), \widetilde{\sigma}(i,j)} \delta_{ij} \right](\mathcal{B} _{kl})_{ij}\\
	&= &\sum_{k,j=1}^n \left[\sum_{i<j}^n \frac{1}{2}A_{\widetilde{\sigma}(k,l), \sigma(j,i)} (\mathcal{B} _{kl})_{ji}+ \sum_{i=j=1}^nA_{\widetilde{\sigma}(k,l), \sigma(i,j)} (\mathcal{B} _{kl})_{ij}+  \frac{1}{2}\sum_{i>j}^nA_{\widetilde{\sigma}(k,l), \sigma(i,j)} (\mathcal{B} _{kl})_{ij}\right]\\
		&= &\sum_{k,j=1}^n \sum_{i\geq j}^n A_{\widetilde{\sigma}(k,l), \sigma(i,j)} (\mathcal{B} _{kl})_{ji}\\
		&= &\sum_{i\geq j}^n \left[\sum_{k<l}^nA_{\sigma(l,k), \sigma(j,i)} (\mathcal{B} _{lk})_{ji}+ \sum_{k=l=1}^nA_{\sigma(k,l), \sigma(i,j)} (\mathcal{B} _{kl})_{ij}+  \sum_{l<k}^nA_{\sigma(k,l), \sigma(i,j)} (\mathcal{B} _{kl})_{ij}\right]\\
		&= &\sum_{i\geq j}^n \left[\sum_{k\geq l}^n A_{\sigma(k,l), \sigma(i,j)} (\mathcal{B} _{kl})_{ij}- \sum_{k=l=1}^nA_{\sigma(k,l), \sigma(i,j)} (\mathcal{B} _{kl})_{ij} \delta_{kl} \right]\\
		&= &\sum_{p,q=1} ^N\left[ 2A_{p,q}B_{p,q}-A_{p,q}B_{p,q}\delta_{{\rm pr}_1(\sigma ^{-1}(p)),{\rm pr}_2(\sigma ^{-1}(p))} \right]= \operatorname{trace}(2AB ^T-A \widetilde{B} ^T)=\langle A, 2B- \widetilde{B}\rangle,
\end{eqnarray*} 
which proves the statement. We emphasize that in the fourth and sixth equalities we used the $n$-symmetry of $\mathcal{B}$. The equality~(\ref{sigma inverse expression}) is proved in a straightforward manner by verifying that $\widetilde{\Sigma}^{-1} \circ \Sigma= \mathbb{I}_{\mathbb{M}_N}$ and $\Sigma \circ \widetilde{\Sigma}^{-1}= \mathbb{I}_{\mathbb{S}_{n ^2}^n}$ using the defining expressions~(\ref{sigma operator}) and~(\ref{sigma inverse expression}).\quad $\blacksquare$

\subsection{Proof of Proposition~\ref{positivity constraint}}

Using the property of the operator $\Sigma$ stated in Proposition~\ref{property of sigmaa statement}, the second equality in~(\ref{vec11 model}) can be rewritten as:
\begin{eqnarray*}
{\rm vech}(H _t)&= & {\rm vech}({\rm math}( {\bf c}))+A {\rm vech}({\bf z}_{t-1}{\bf z}_{t-1}^T)+B {\rm vech}(H_{t-1})\\
	&= &{\rm vech}({\rm math}( {\bf c})) + {\rm vech}(\Sigma (A)\bullet ({\bf z}_{t-1}{\bf z}_{t-1}^T))+ {\rm vech}(\Sigma(B)\bullet H_{t-1}),
\end{eqnarray*}
or, equivalently:
\begin{equation*}
H _t={\rm math}( {\bf c})+\Sigma (A)\bullet ({\bf z}_{t-1}{\bf z}_{t-1}^T)+ \Sigma(B)\bullet H_{t-1}.
\end{equation*}
In view of this expression and in the terms of the statement of the proposition, it suffices to show that both $\Sigma (A)\bullet ({\bf z}_{t-1}{\bf z}_{t-1}^T)$ and $\Sigma(B)\bullet H_{t-1}$ are positive semidefinite provided that $H_{t-1}$ is positive semidefinite. Regarding $\Sigma (A)\bullet ({\bf z}_{t-1}{\bf z}_{t-1}^T)$, consider $\mathbf{v}\in \mathbb{R}^{n ^2} $. Then
\begin{eqnarray*}
\langle\mathbf{v}, \Sigma (A)\bullet ({\bf z}_{t-1}{\bf z}_{t-1}^T) \mathbf{v}\rangle&= &
\sum_{i,j=1}^{n ^2} v _i (\Sigma (A)\bullet ({\bf z}_{t-1}{\bf z}_{t-1}^T))_{ij}v _j
=\sum_{i,j=1}^{n ^2} v _i \operatorname{trace}(\Sigma (A)_{ij}({\bf z}_{t-1}{\bf z}_{t-1}^T))v _j\\
	&= &\sum_{i,j=1}^{n ^2} v _i \operatorname{trace}({\bf z}_{t-1}^T\Sigma (A)_{ij}{\bf z}_{t-1})v _j
	=\sum_{i,j,k,l=1}^{n ^2} v _i z_{t-1,k}^T(\Sigma (A)_{ij})_{kl}z_{t-1,l}v _j\\
	&= & \langle\mathbf{v}\otimes {\bf z}_{t-1}, \Sigma(A)(\mathbf{v}\otimes {\bf z}_{t-1})\rangle,
\end{eqnarray*}
which is greater or equal to zero due to the positive semidefiniteness hypothesis on $\Sigma(A)$. In the last equality we used~(\ref{components block matrices}).

As to $\Sigma(B)\bullet H_{t-1}$, we start by noticing that  $H_{t-1}=E_{t-1}[{\bf z}_{t-1}{\bf z}_{t-1}^T]$ and hence $\Sigma(B)\bullet H_{t-1}=\Sigma(B)\bullet E_{t-1}[{\bf z}_{t-1}{\bf z}_{t-1}^T]$. This equality, as well as the linearity of the conditional expectation allows us to use virtually the same argument as above. Indeed, for any $\mathbf{v}\in \mathbb{R}^{n ^2} $
\begin{eqnarray*}
\langle\mathbf{v},\Sigma(B)\bullet H_{t-1} \mathbf{v}\rangle&= &
\sum_{i,j=1}^{n ^2} v _i \operatorname{trace}(\Sigma (B)_{ij}E_{t-1}[{\bf z}_{t-1}{\bf z}_{t-1}^T])v _j=\sum_{i,j=1}^{n ^2} E_{t-1}[v _i \operatorname{trace}(\Sigma (B)_{ij}{\bf z}_{t-1}{\bf z}_{t-1}^T)v _j]\\
	&= & E_{t-1}[\langle\mathbf{v}\otimes {\bf z}_{t-1}, \Sigma(B)(\mathbf{v}\otimes {\bf z}_{t-1})\rangle],
\end{eqnarray*}
which is greater or equal to zero due to the positive semidefiniteness hypothesis on $\Sigma(B)$. \quad $\blacksquare$

\subsection{Proof of Proposition~\ref{Second order stationarity constraints}}

We start by noticing that the VEC(1,1) model is by construction a white noise and hence it suffices to establish the stationarity of the variance. Indeed, for any $t,h \in \mathbb{N}$ we compute the autocovariance function $\Gamma$:
\begin{multline}
\label{model is white noise}
\Gamma(t,t+h):=E \left[{\bf z}_t{\bf z}_{t+h}^T\right]=E\left[ E _t\left[H_t^{1/2}\boldsymbol{\epsilon _t}\boldsymbol{\epsilon _{t+h}}H_{t+h}^{1/2}\right]\right]\\
=E\left[H_t^{1/2} E _t\left[\boldsymbol{\epsilon _t}\boldsymbol{\epsilon _{t+h}}\right]H_{t+h}^{1/2}\right]=\delta_{h0}E\left[H_t^{1/2} H_{t+h}^{1/2}\right].
\end{multline}
Consequently, we just need to prove the existence of a solution for which $\Gamma(t,t)=E \left[H _t\right] $ or, equivalently  $E[ {\bf h}_t] $, is time independent. We first notice that 
\begin{equation*}
E[ {\bf h}_t]=E \left[{\bf c}+A \boldsymbol{\eta}_{t-1}+B  \boldsymbol{h}_{t-1}\right]=E \left[{\bf c}+A \boldsymbol{h}_{t-1}+B  \boldsymbol{h}_{t-1}\right]+A \,E \left[\boldsymbol{\eta}_{t-1}-\boldsymbol{h}_{t-1}\right]=E \left[{\bf c}+A \boldsymbol{h}_{t-1}+B  \boldsymbol{h}_{t-1}\right],
\end{equation*} 
since $A \,E \left[\boldsymbol{\eta}_{t-1}-\boldsymbol{h}_{t-1}\right]=0 $ by~(\ref{model is white noise}). Now, for any $k>0$
\begin{equation*}
E[ {\bf h}_t]={\bf c}+(A+B)E \left[{\bf h}_{t-1}\right]=\sum_{j=0}^k (A+B)^j{\bf c}+(A+B)^{k+1}E \left[{\bf h}_{t-k-1}\right].
\end{equation*}
If all the eigenvalues of $A+B $ are smaller than one in modulus then (see, for example~\cite[Appendix A.9.1]{luetkepohl:book})
\begin{equation*}
\sum_{j=0}^k (A+B)^j{\bf c}\xrightarrow[k \rightarrow \infty]{}(\mathbb{I}_N-A-B) ^{-1}{\bf c}, \qquad  \mbox{and}  
\qquad
(A+B)^{k+1}E \left[{\bf h}_{t-k-1}\right]\xrightarrow[k \rightarrow \infty]{} 0,
\end{equation*}
in which case $E[ {\bf h}_t] $ is time independent and
\begin{equation*}
\Gamma(0)={\rm math}(E[ {\bf h}_t])={\rm math}((\mathbb{I}_N-A-B) ^{-1}{\bf c}).
\end{equation*}
The sufficient condition in terms of the top singular value $\sigma_{{\rm max}}(A+B)$ of $A+B $  is a consequence of the fact that (see for instance~\cite[Theorem 5.6.9]{Horn:Johnson}) $| \lambda(A+B)|\leq \sigma_{{\rm max}}(A+B) $, for any eigenvalue  $\lambda(A+B) $ of $A+B $. \quad $\blacksquare$

\subsection{Proof of Proposition~\ref{gradient by recursion}}

The chain rule implies that for any perturbation $\Delta $ in the $\theta$ direction
\begin{equation*}
d_\theta l _t \cdot  \Delta= d_{H _t} l _t(H _t(\theta))\cdot T _\theta H _t \cdot \Delta=\langle \nabla_{H _t} l _t, T_{\theta}H _t \cdot  \Delta\rangle=\langle T_{\theta}^\ast H _t \cdot\nabla_{H _t} l _t,   \Delta\rangle,
\end{equation*}
which proves that $ \nabla _{ \theta } l _t=  T^\ast _{ \theta }  H _t \cdot  \nabla_{H _t} l _t $  and hence~(\ref{general expression of gradient}) follows. We now establish~(\ref{gradient with respect to Ht}) by showing separately that
\begin{equation}
\label{two separate gradients}
\nabla_{H _t} \log(\det(H _t))= H _t^{-1} \quad \mbox{and} \quad \nabla_{H _t} \left(- \frac{1}{2} {\bf z} _t^T H _t ^{-1} {\bf z} _t\right)= \frac{1}{2} \left(H _t ^{-1} {\bf z} _t{\bf z} _t^T H _t ^{-1}  \right).
\end{equation}
In order to prove the first expression we start by using the positive semidefinite character of $H _t  $ in order to write $H _t=V D V ^T $. $V$  is an orthogonal matrix and $D $ is diagonal with non-negative entries; it has hence a unique square root  $D ^{1/2} $ that we can use to write $H _t=V D V ^T =(V D ^{1/2})(V D ^{1/2})^T$. Let  $\delta \in \mathbb{R} $ and $\Delta\in \Bbb S _n$. We have
\begin{eqnarray*}
 \log(\det(H _t+ \delta \Delta)) &= &\log(\det((V D ^{1/2})(V D ^{1/2})^T+ \delta \Delta))\\
 	&=&\log(\det((V D ^{1/2})(\mathbb{I}_n+\delta(D^{-1/2}V ^T)\Delta (VD^{-1/2}))(V D ^{1/2})^T))\\
	&= &\log(\det(V D ^{1/2})\det(\mathbb{I}_n+\delta(D^{-1/2}V ^T)\Delta (VD^{-1/2}))\det(V D ^{1/2})^T)\\
	&= &\log(\det((V D ^{1/2})(V D ^{1/2})^T)\det(\mathbb{I}_n+\delta(D^{-1/2}V ^T)\Delta (VD^{-1/2})))\\
	&=&\log(\det(H _t)\det(\mathbb{I}_n+\delta\Xi)),
\end{eqnarray*}
with $ \Xi:=(D^{-1/2}V ^T)\Delta (VD^{-1/2}) $. This matrix is symmetric and hence normal and diagonalizable; let $\left\{ \lambda _1, \ldots, \lambda _n \right\} $ be its eigenvalues. We hence have that
\begin{eqnarray*}
dH _t \cdot \Delta &= &\left.\frac{d}{d\delta}\right|_{\delta=0} \log(\det(H _t+ \delta \Delta))=   \left.\frac{d}{d\delta}\right|_{\delta=0} \log(\det(H _t))+\log\left(\prod_{i=1}^n(1+ \delta \lambda _i)\right)=\left.\frac{d}{d\delta}\right|_{\delta=0}\sum_{i=1}^n \log(1+ \delta \lambda _i)\\
	&=&\sum_{i=1}^n \lambda _i= \operatorname{trace}((D^{-1/2}V ^T)\Delta (VD^{-1/2}) )= \operatorname{trace}((VD^{-1/2})(D^{-1/2}V ^T) \Delta)= \operatorname{trace}(H _t ^{-1} \Delta),
\end{eqnarray*}
which proves $\nabla_{H _t} \log(\det(H _t))= H _t^{-1} $. Regarding the second expression in~(\ref{two separate gradients}) we define $f(H _t):= - \frac{1}{2} {\bf z} _t^T H _t ^{-1} {\bf z} _t $ and note that
\begin{eqnarray*}
d f (H _t)\cdot \Delta&=& \left.\frac{d}{dt}\right|_{t=0} - \frac{1}{2} {\bf z} _t^T (H _t+t \Delta) ^{-1} {\bf z} _t= \left.\frac{d}{dt}\right|_{t=0} - \frac{1}{2} {\bf z} _t^T (\mathbb{I}_n+t H _t ^{-1}\Delta) ^{-1} H _t ^{-1}{\bf z} _t\\
	&= &\left.\frac{d}{dt}\right|_{t=0} - \frac{1}{2} {\bf z} _t^T (\mathbb{I}_n+t H _t ^{-1}\Delta) ^{-1} H _t ^{-1}{\bf z} _t=\left.\frac{d}{dt}\right|_{t=0} - \frac{1}{2} \sum_{k=0}^{\infty}(-1) ^{k}t^k{\bf z} _t^T (H _t ^{-1} \Delta)^k H _t ^{-1}{\bf z} _t\\
	&=&\frac{1}{2}{\bf z} _t^T H _t ^{-1} \Delta H _t ^{-1}{\bf z} _t= \frac{1}{2} \operatorname{trace}(H _t ^{-1}{\bf z}_t {\bf z} _t ^T H _t ^{-1} \Delta),
\end{eqnarray*}
which implies that $\nabla _{H _t }f=\frac{1}{2}\left(H _t ^{-1}{\bf z}_t {\bf z} _t ^T H _t ^{-1}  \right)$, as required.

In order to prove~(\ref{tht1})--(\ref{tht3}) we notice that  the second equation in~(\ref{vec11 model}) can be rewritten using the vech and math operators as
\begin{equation}
\label{ht vs Ht}
H _t= {\rm math}\left({\bf c}+A \boldsymbol{\eta}_{t-1}+B {\rm vech}(H_{t-1})\right).
\end{equation}
We now show~(\ref{tht1}). Let $\mathbf{v} \in \mathbb{R}^N  $ and $\Delta \in \Bbb S _n $  arbitrary. Identity~(\ref{ht vs Ht}) and the linearity of the various mappings involved imply that
$
T_{{\bf c}}H _t \cdot \mathbf{v}= {\rm math}\left( \mathbf{v}+B {\rm vech}(T_{{\bf c}}H_{t-1}\cdot \mathbf{v})\right)
$
and hence
\begin{eqnarray*}
\langle T^\ast _{{\bf c}}H _t \cdot\Delta,  \mathbf{v}\rangle &= & \langle\Delta, T_{{\bf c}}H _t \cdot \mathbf{v}\rangle= \langle \Delta, {\rm math}\left({\bf v}+B {\rm vech}(T_{{\bf c}}H_{t-1}\cdot \mathbf{v})\right)\rangle\\
	&= &\langle{\rm math}^\ast(\Delta)+T _{{\bf c}}^\ast H_{t-1}\cdot {\rm vech}^\ast(B ^T{\rm math}^\ast(\Delta) ),\mathbf{v}\rangle.
\end{eqnarray*}
The proof of~(\ref{tht2}) follows a similar scheme. By~(\ref{ht vs Ht}) we have that for any $M \in \mathbb{M}_N$:
\begin{equation}
\label{a differential}
T_{A}H _t \cdot M= {\rm math}\left(M \boldsymbol{\eta}_{t-1}+B {\rm vech}(T_{A}H_{t-1}\cdot M)\right).
\end{equation}
Consequently, for any $\Delta \in \Bbb S _n $
\begin{eqnarray*}
\langle T _A^\ast H _t \cdot \Delta, M\rangle&=&\langle \Delta, T _A H _t\rangle=\langle \Delta, {\rm math}\left(M \boldsymbol{\eta}_{t-1}+B {\rm vech}(T_{A}H_{t-1}\cdot M)\right)\rangle\\
	&=& \langle{\rm math}^\ast(\Delta)\cdot \boldsymbol{\eta}_{t-1}^T+T _A^\ast H_{t-1}\cdot {\rm vech}^\ast(B ^T{\rm math}^\ast(\Delta) ), \Delta\rangle.
\end{eqnarray*}
Finally,~(\ref{tht3}) is proved analogously replacing~(\ref{a differential}) by its $B$ counterpart, namely, $$T_{B}H _t \cdot M= {\rm math}\left(M {\rm vech}(H_{t-1})+B {\rm vech}(T_{B}H_{t-1}\cdot M)\right). \quad \blacksquare$$

\subsection{Proof of Proposition~\ref{estimate number iterations}}

An inductive argument using~(\ref{tht1})--(\ref{tht3}) guarantees that for any  $t ,k \in \mathbb{N} $, $k \leq t $
\begin{eqnarray}
T_{{\bf c}}H _t^\ast  \cdot \Delta &= &\sum_{i=1}^k B^{i-1\,T} {\rm math}^\ast (\Delta)+T _{{\bf c}}^\ast H_{t-k}\cdot {\rm vech}^\ast(B ^{k\,T}{\rm math}^\ast(\Delta) ), \label{thtrec1}\\  
T _A^\ast H _t \cdot \Delta&=& \sum_{i=1}^k B^{i-1\,T} {\rm math}^\ast(\Delta)\cdot \boldsymbol{\eta}_{t-i}^T+T _A^\ast H_{t-k}\cdot {\rm vech}^\ast(B ^{k\,T}{\rm math}^\ast(\Delta) ),\label{thtrec2}\\  
T _B^\ast H _t \cdot \Delta&=& \sum_{i=1}^k B^{i-1\,T} {\rm math}^\ast(\Delta)\cdot {\rm vech}(H_{t-i})^T+T _B^\ast H_{t-k}\cdot {\rm vech}^\ast(B ^{k\,T}{\rm math}^\ast(\Delta) ),\label{thtrec3}
\end{eqnarray}
The first expression with $k=t $  and the norm estimate~(\ref{operator norm math3}) imply that
\begin{equation}
\label{tstarforc}
\|T_{{\bf c}}H _t^\ast  \cdot \Delta\|=\left\|\sum_{i=1}^t B^{i-1\,T} {\rm math}^\ast (\Delta)\right\|\leq \sqrt{2}\sum_{i=1}^t \|B\|_{{\rm op}}^{i-1}\| \Delta\|\leq \frac{\sqrt{2}\| \Delta\|}{1-\|B\|_{{\rm op}}}.
\end{equation}
We now use~(\ref{thtrec1}) for an arbitrary $k$ as well as~(\ref{operator norm math2}) and~(\ref{tstarforc}) and write
\begin{multline}
\|(T^\ast _{{\bf c} } H _t -T^\ast _{ {\bf c} } H _t ^k) \cdot \Delta\|=\|T _{{\bf c}}^\ast H_{t-k}\cdot {\rm vech}^\ast(B ^{k\,T}{\rm math}^\ast(\Delta) ) \|\\
\leq \| T _{{\bf c}}^\ast H_{t-k} \|_{{\rm op}}\|{\rm vech}^\ast\|_{{\rm op}}\|B\|_{{\rm op}}^k\|{\rm math}^\ast\|_{{\rm op}}\| \Delta\|\leq \frac{2\| \Delta\|\|B\|_{{\rm op}}^k}{1-\|B\|_{{\rm op}}}.
\end{multline}
The computability constraint {\bf (CC)} implies that $\|B\|_{{\rm op}}\leq 1- \widetilde{\epsilon}_{B} $ and hence $\|T^\ast _{{\bf c} } H _t -T^\ast _{ {\bf c} } H _t ^k\|_{{\rm op}}\leq 2(1- \widetilde{\epsilon } _B)^k/\widetilde{\epsilon } _B$. A straightforward computation shows that if we want this upper bound for the error to be smaller than a certain $\delta>0$, that is $2(1- \widetilde{\epsilon } _B)^k/\widetilde{\epsilon } _B < \delta $ then it suffices to take
\begin{equation}
\label{first k estimation}
k>\frac{\log \left(\frac{\widetilde{\epsilon} _B\delta}{2}\right)}{\log (1- \widetilde{\epsilon} _B)}.
\end{equation}
We now tackle the estimation of the truncation error in mean in the $A$ variable. Firstly, we recall that by~(\ref{model is white noise}) and in the presence of the stationarity constraint  $E[\boldsymbol{\eta} _t]=E[{\bf h}_t ]=(\mathbb{I}_N-A-B)^{-1} {\bf c}$. The first consequence of this identity is that if we take the expectations of both~(\ref{thtrec2}) and~(\ref{thtrec3}) we see that $\|E\left[T^\ast _{A } H _t \cdot \Delta\right]\| $  and  $\|E\left[T^\ast _{B } H _t \cdot \Delta\right]\| $ are determined by exactly the same recursions and hence the error estimations for both variables are going to be the same. Also, by~(\ref{thtrec2})
\begin{multline}
\label{byafora}
\|E\left[T^\ast _{A } H _t \cdot \Delta\right]\|=\left\|\sum_{i=1}^t B^{i-1\,T} {\rm math}^\ast (\Delta)\cdot  E[\boldsymbol{\eta}_{t-i}^T]\right\|\leq \sqrt{2}\| \Delta\|\|E[{\bf h}_t]\|\sum_{i=1}^t \|B\|_{{\rm op}}^{i-1}\\
\leq \sqrt{2}\| \Delta\|\|(\mathbb{I}_N-A-B) ^{-1}{\bf c}\|/ \widetilde{\epsilon}_B \leq \sqrt{2}\| \Delta\|\| {\bf c}\|/ \epsilon_{AB}\widetilde{\epsilon}_B. 
\end{multline}
The last inequality is a consequence of the constraints {\bf (SC)} and {\bf (PC)}. Indeed,
\begin{equation*}
\|(\mathbb{I}_N-A-B) ^{-1}{\bf c}\|=\left\|\sum_{i=0}^{\infty} (A+B)^i{\bf c}\right\|\leq\sum_{i=0}^{\infty} \left\|(A+B)\right\|_{{\rm op}}^i\|{\bf c}\|\leq \sum_{i=0}^{\infty} (1- \epsilon_{AB})^i\|{\bf c}\|= \frac{\| {\bf c}\|}{\epsilon_{AB}}.
\end{equation*}
Now, by~(\ref{thtrec2}) and~(\ref{byafora}), 
\begin{multline}
\|E\left[(T^\ast _{A} H _t -T^\ast _{A} H _t ^k) \cdot \Delta\right]\|=\|E\left[T _{A}^\ast H_{t-k}\cdot {\rm vech}^\ast(B ^{k\,T}{\rm math}^\ast(\Delta) )\right] \|\\
\leq \| T _{A}^\ast H_{t-k} \|_{{\rm op}}\|{\rm vech}^\ast\|_{{\rm op}}\|B\|_{{\rm op}}^k\|{\rm math}^\ast\|_{{\rm op}}\| \Delta\|\leq \frac{2\| \Delta\|\| {\bf c}\|}{\epsilon_{AB}\widetilde{\epsilon} _B}(1-\widetilde{\epsilon} _B)^k,
\end{multline}
which proves~(\ref{ub2}). If we want this upper bound for the error to be smaller than a certain $\delta>0$, we have to make the number of iterations $k$ big enough so that 
\begin{equation*}
\frac{2\| {\bf c}\|}{\epsilon_{AB}\widetilde{\epsilon} _B}(1-\widetilde{\epsilon} _B)^k< \delta \quad \mbox{that is} \quad (1-\widetilde{\epsilon} _B)^k= \frac{\delta \epsilon_{AB}\widetilde{\epsilon}_B}{2 \| {\bf c}\|}\leq \frac{\delta \epsilon_{AB}\widetilde{\epsilon}_B}{2 \epsilon_{{\bf c}}}.
\end{equation*}
This relation, together with~(\ref{first k estimation}) proves the estimate~(\ref{k estimate}). \quad $\blacksquare$

\addcontentsline{toc}{section}{Bibliography}
\bibliographystyle{alpha}
\bibliography{/Users/JP17/JPO_synch/BiblioData/bibliography_econometry}
\footnotesize

\end{document}